\documentclass[aps, showpacs, pra, superscriptdaddress,   % notitlepage, 
eqsecnum, twocolumn] {revtex4-1}
\usepackage{hyperref}
\usepackage{amsmath }
\usepackage{amssymb}
\usepackage{epsfig}
\usepackage{natbib}
\newcommand*{\be}{\begin{equation}}
\newcommand*{\ee}{\end{equation}}
\newcommand*{\bea}{\begin{eqnarray}}
\newcommand*{\eea}{\end{eqnarray}}

\newcommand{\sech}{\, \mathrm{sech}}

\begin{document}

\title[]{Optical solitons in $\mathcal{PT}$-symmetric nonlinear couplers with gain and loss}

\author{N V Alexeeva} 
  \altaffiliation{On sabbatical leave from the Department of Mathematics, University of Cape Town, Rondebosch 7701, South Africa}

 \author{I  V  Barashenkov}
 \altaffiliation{On sabbatical leave from the Department of Mathematics, University of Cape Town, South Africa;
 also at the Joint Institute for Nuclear Research,  Dubna, Russia}
% \altaffiliation{On sabbatical leave from the Department of Mathematics, University of Cape Town, Rondebosch 7701, South Africa}
%\thanks{also at the Joint Institute for Nuclear Research, Dubna, Russia}

 \author{Andrey A Sukhorukov}

 \author{Yuri S Kivshar}
 \affiliation{Nonlinear Physics Centre, Australian National University, Canberra ACT 0200,  Australia}

\begin{abstract}
We study spatial and temporal solitons in the
 $\mathcal{PT}$ symmetric coupler with gain in one waveguide and loss in the other.
Stability properties of
the high- and  low-frequency solitons are found to be completely determined by a single combination
 of the soliton's amplitude and
 the gain/loss coefficient of the waveguides. 
The unstable perturbations of the high-frequency soliton
break the symmetry between its active and lossy components
which  results in a blowup of the soliton or a formation of a long-lived breather state.
The unstable perturbations
of the low-frequency soliton
 separate its two components  in space
blocking the power drainage of  the active component and cutting the power supply to the  lossy one.
Eventually this  also leads to the blowup or  breathing.

\end{abstract}

\maketitle

\section{Introduction}

Optical solitons are formed when nonlinear effects compensate the diffractive broadening of light beams (spatial solitons) or dispersive spreading of optical pulses (temporal solitons).                                  
% Solitons can even exist both in conservative regimes and in structures containing gain and loss regions. 
Although these localized structures  arise both in conservative settings and in systems with active and lossy elements,
%The properties of dissipative optical solitons in structures with gain and loss~\cite{Rosanov:2002:SpatialHysteresis, Akhmediev:2005:DissipativeSolitons} tend to differ fundamentally from their counterparts %     
%  in conservative structures~\cite{Kivshar:2003:OpticalSolitons}. 
properties of dissipative optical solitons~\cite{Rosanov:2002:SpatialHysteresis, Akhmediev:2005:DissipativeSolitons} 
 show significant differences from those of their conservative counterparts~\cite{Kivshar:2003:OpticalSolitons}. 
In particular,  the amplitudes of solitons in conservative systems are free to vary over continuous ranges
whereas the generic dissipative systems can only support 
solitons at special amplitude values determined by the balance between gain and loss.

In an interesting turn of events, it was realised recently that there is a class of optical systems where
  dissipative solitons arise in continuous families. These 
   systems consist of  elements with  gain and loss  arranged in 
    a particular symmetric way~\cite{Musslimani:2008-30402:PRL}.  The symmetry here can be interpreted 
   as an optics equivalent ~\cite{Ruschhaupt:2005-L171:JPA, El-Ganainy:2007-2632:OL}
   of the ${\cal PT}$ (parity-time) symmetry in 
    quantum mechanics~\cite{Bender:1998-5243:PRL, Bender:2002-270401:PRL, Bender:2003-1095:AMJP, Bender:2007-947:RPP}. 
 % However in recent years it was revealed that dissipative solitons can form continuous families in optical structures where the distribution of gain and loss regions conforms to parity-time (${\cal PT}$) symmetry condition~\cite{Musslimani:2008-30402:PRL}, which can be formulated for optical structures~\cite{Ruschhaupt:2005-L171:JPA, El-Ganainy:2007-2632:OL}
   % A distinguishing feature of such potentials is that their spectrum can be real, meaning that all linear eigenmodes have non-decaying and non-growing amplitudes due to an effective cancelation between gain and loss. Nevertheless, interference of several modes gives rise to non-conservative evolution, and a range of phenomena have been predicted which are impossible in conservative systems~\cite{Berry:2008-244007:JPA, Makris:2008-103904:PRL, Longhi:2009-123601:PRL,
% Bendix:2009-30402:PRL, West:2010-54102:PRL, Longhi:2010-22102:PRA}.
The ${\cal PT}$-symmetric potentials in quantum mechanics are essentially complex potentials which however exhibit a purely real spectrum of energies,
with the implication that their time-dependent eigenfunctions show no decay or growth.
Despite this similarity to the hermitian quantum mechanics, the $\cal{PT}$-symmetric  quantum systems display a variety of anomalous phenomena stemming from the 
nonhermitian mode interference~\cite{Berry:2008-244007:JPA, Makris:2008-103904:PRL, Longhi:2009-123601:PRL,
 Bendix:2009-30402:PRL, West:2010-54102:PRL, Longhi:2010-22102:PRA}.

The  first experimental demonstrations of the ${\cal PT}$-symmetric effects in optics were in two-waveguide directional linear couplers
composed of waveguides with gain and loss~\cite{Guo:2009-93902:PRL, Ruter:2010-192:NPHYS}. 
% Theoretical studies predict applications of such couplers composed of waveguides with gain and loss for all-optical signal control in the nonlinear regime~\cite{Chen:1992-239:IQE, Ramezani:2010-43803:PRA, Sukhorukov:2010-43818:PRA, Dmitriev:2011-13833:PRA}. Arrays of such ${\cal PT}$-symmetric couplers have been suggested as a feasible platform for control over spatial beam dynamics including formation and switching of spatial solitons~\cite{Dmitriev:2010-2976:OL, Zheng:2010-10103:PRA, Suchkov:2011-46609:PRE}.
Theoretical analyses suggest that  such couplers, operating in the nonlinear regime, can be used
for the all-optical signal control~\cite{Chen:1992-239:IQE, Ramezani:2010-43803:PRA, Sukhorukov:2010-43818:PRA, Dmitriev:2011-13833:PRA}. 
Arrays of the ${\cal PT}$-symmetric couplers were proposed as a feasible means of control of the spatial beam dynamics, including 
the formation and switching of spatial solitons~\cite{Dmitriev:2010-2976:OL, Zheng:2010-10103:PRA, Suchkov:2011-46609:PRE}.

\begin{figure}[t]
 \begin{center} 
 \includegraphics*[width=\linewidth]{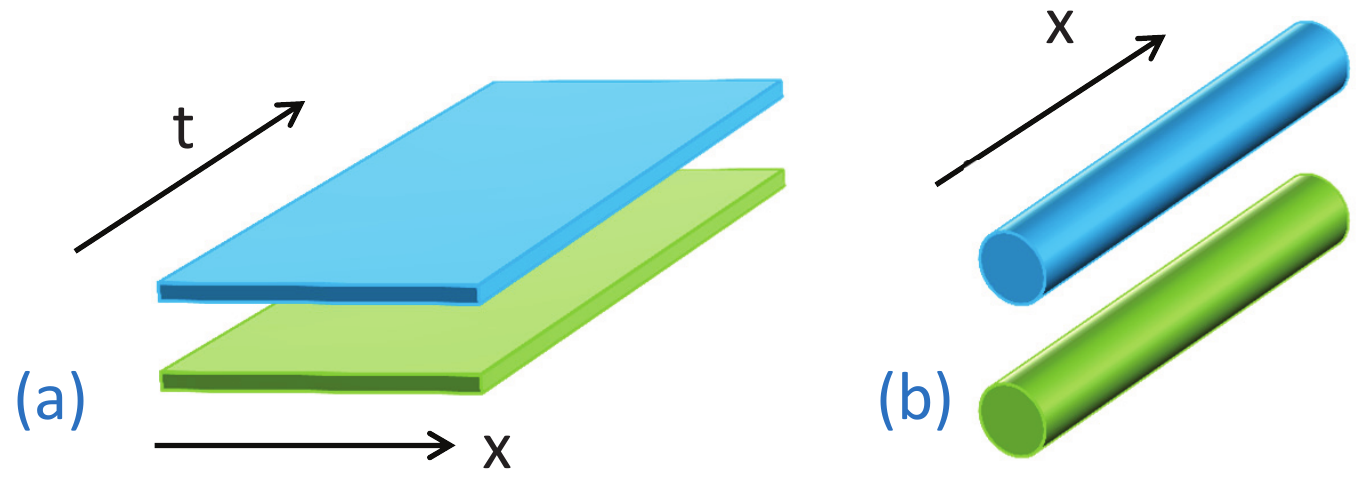} % {PTsketch.eps} %  {imag11.eps}
  \caption{A schematic of ${\cal PT}$-symmetric coupled waveguides with gain (top waveguide) and loss (bottom waveguide).
(a)~Two waveguides on the plane where $t$ denotes the logitudinal and $x$  the transversal spatial coordinate.
The transverse $x$-profiles of the stationary light beams evolve
as  they extend in the $t$ direction.
(b)~A pair of one-dimensional waveguides  where light pulses  undergo temporal evolution as they travel along the $x$ axis.   }
 \label{scheme}
 \end{center}
 \end{figure}

This paper is concerned  with  the ${\cal PT}$-symmetric couplers with 
an extra spatial or temporal degree of freedom. We  consider the situation of stationary light beams in the  coupled planar waveguides 
[i.e. waveguides extended in the transverse direction, Fig.~\ref{scheme}(a)] 
and that of the optical pulses in coupled one-dimensional 
waveguides [Fig.~\ref{scheme}(b)].  The configuration shown in Fig.~\ref{scheme}(a) can also be seen as  
 the strong-coupling limit of  the array of coupled dimers discussed in the recent Ref.~\cite{Suchkov:2011-46609:PRE}.

Our study focuses on spatial and temporal solitons in such couplers.
The reader should be alerted up front  that we use
the term ``soliton"
 simply as a synonym for  solitary wave or localised pulse. 
No {\it a~priori\/} stability is implied by the use of this term.
It is the objective of this study to 
classify 
  the ${\cal PT}$-symmetric solitons  into
stable and unstable ones.

In addition to the analysis of the solitons'  stability
properties  and numerical study of the  linearisation eigenvalues,
we  uncover the instability mechanisms
 and  simulate the nonlinear evolution of the unstable solitons.

The paper is organized as follows. 
In the next section
(Sec.~\ref{Model})  we formulate the mathematical model and identify  physically
meaningful integral characteristics of the associated evolution. 
Two families of high- and low-frequency bright soliton solutions of the model
are introduced in Sec.~\ref{Solitons}. In the following section, Sec.~\ref{Stability},  we outline the general framework for their
stability analysis. The stability eigenvalues of the high-frequency soliton are classified
in section \ref{High}; there, we also follow the nonlinear evolution of instability when the soliton is found to be unstable.
Section \ref{Low} contains a similar study of the low-frequency soliton.
Finally, section \ref{Conclusions} summarises conclusions of our work
while three Appendices detail mathematical analyses of the stability eigenvalues.

%-----------------------------------------------------------------
\section{The model} 
\label{Model}
%-----------------------------------------------------------------

To describe the dynamics of stationary light beams and pulses in coupled waveguides illustrated in Fig.~\ref{scheme}, we extend 
the equations of the nonlinear $\mathcal{PT}$-symmetric coupler~\cite{Chen:1992-239:IQE, Ramezani:2010-43803:PRA, Sukhorukov:2010-43818:PRA}.
In the physical setting of Fig.~\ref{scheme}(a),  our extension takes into account the diffraction of the beam
while in the situation represented by Fig.~\ref{scheme}(b),  we extend the system to include the effect of the pulse dispersion.
The resulting equations have the following dimensionless form:
%
%\begin{subequations}
\begin{align}
i u_t + u_{xx} + 2 |u|^2 u= -v + i \gamma u, \nonumber \\
iv_t+  v_{xx} + 2|v|^2v = -u - i \gamma v. \label{A1}
\end{align}
%\end{subequations}
%
We note that this model was originally introduced as the continuous limit for a one-dimensional  array of 
$\cal{PT}$-symmetric coupled waveguides~\cite{Suchkov:2011-46609:PRE}.

In Eqs.\eqref{A1},  the  $u$ and $v$ variables are the normalized complex mode amplitudes in the top and bottom waveguides  of  Fig.~\ref{scheme}.
Considering stationary beams in the planar geometry of Fig.~\ref{scheme}(a), 
the        $t$       variable is the (spatial) coordinate in the propagation direction
while $x$ is the transversal coordinate.  In the temporal pulse interpretation [Fig.~\ref{scheme}(b)], 
$t$ stands for time and $x$ for the spatial coordinate in the frame moving with the pulse group velocity.
 Here we assume that the group velocities and second-order dispersions in the  waveguides are matched so as to satisfy the $\mathcal{PT}$-symmetry condition.
By scaling the $x$ variable properly,
we have normalized  the coefficients in front of $u_{xx}$  and $v_{xx}$ terms
to unity. (These terms account for the diffraction of spatial beams and dispersion of temporal pulses.)

We assume that the Kerr nonlinearity coefficients
in the two waveguides (coefficients in front of  $|u|^2 u$ and $|v|^2v$)
 have the same value as this is necessary for the existence of solitons~\cite{Suchkov:2011-46609:PRE}. 
  (These coefficients have been normalized to $2$ by scaling the mode amplitudes.)
  To ensure the existence of  bright solitons, we take equal
  signs in front of the diffraction/dispersion and nonlinear terms.
   This corresponds to the self-focusing nonlinearity in the case of  beams [Fig.~\ref{scheme}(a)], 
   % while in the case of  pulses, this choice pertains to 
   and to 
    the anomalous dispersion with positive Kerr nonlinearity or normal dispersion with negative nonlinearity
    --- in the case of pulses [Fig.~\ref{scheme}(b)].

Whether Eqs.\eqref{A1} are employed to describe  the stationary planar beams or temporal pulses, 
the first terms in the right-hand sides of \eqref{A1}
account for  the coupling between the modes propagating in 
 the two waveguides. The $\gamma$-terms describe the gain in one and loss in the other waveguide.
 Without loss of generality $\gamma$ can be taken positive; this choice corresponds to the gain 
  in the top  and loss in the bottom waveguide.
 The gain and loss coefficients are taken equal
  to conform to the  $\mathcal{PT}$-symmetry condition~\cite{Ruter:2010-192:NPHYS}.

We close this section by noting 
several physically meaningful quantities which  prove useful in the understanding of the dynamics
described by Eqs.\eqref{A1}.
The first pair of variables give the powers associated with the $u$ and $v$ fields, respectively:
\be
\mathcal P_u=\int |u|^2 dx, \quad \mathcal P_v = \int |v|^2 dx.
\label{PP}
\ee
Neither individual powers nor their sum are conserved
if $\gamma \neq 0$; however, the rate of change of the total power has a simple and insightful expression:
\be
\frac{d}{dt} (\mathcal{P}_u+\mathcal{P}_v)  = 2 \gamma (\mathcal{P}_u-\mathcal{P}_v).
\label{P} 
\ee

The momenta carried by the $u$ and $v$ components,
\be
 \mathcal{M}_u= \frac{i}{2} \int (u_x^* u-u_xu^*)dx, \quad  \mathcal{M}_v = \frac{i}{2} \int (v_x^* v -v_xv^*)dx,
 \label{MM}
 \ee
are not conserved either. The total momentum satisfies 
\be
\frac{d}{dt} ({\mathcal M}_u +  { \mathcal M}_u)= 2 \gamma ( { \mathcal M}_u   - {\mathcal M}_v).
\label{M}
\ee

 Finally we note the rate equation
 \be \frac{d \mathcal H}{dt}= 2\gamma 
 (\mathcal R_u -\mathcal R_v), \label{H}
 \ee
 where
 \be
 \mathcal H= \int \left[ |u_x|^2+ |v_x|^2-(|u|^4+ |v|^4) -(vu^*+v^*u) \right] dx
 \label{HH}
\ee
and 
\be
\mathcal R_u= \int  (|u_x|^2-2|u|^4) dx,
\quad
\mathcal R_v= \int (|v_x|^2 -2 |v|^4) dx.
\label{FF}
\ee
 The integral $\mathcal H$ plays the role of  the 
 Hamiltonian of Eqs.\eqref{A1} in the situation where there is no loss or gain ($\gamma=0$).

An immediate consequence of Eqs.\eqref{P}, \eqref{M} and \eqref{H} is that all stationary states 
in the system \eqref{A1} have to 
display  symmetry between their two components: $\mathcal{P}_u=\mathcal{P}_v$, 
$\mathcal{M}_u=\mathcal{M}_v$, 
$\mathcal{R}_u=\mathcal{R}_v$.

\section{Solitons}
\label{Solitons}

Proceeding to the analysis of solutions to the system \eqref{A1}, it is convenient to make
a change of variables
\be
u(x,t)=  e^{i(\Omega t- \theta)} U(x,t), \quad
v(x,t)=  e^{i \Omega t} V(x,t),
\label{A40}
\ee
where  $\theta$ is a constant angle satisfying
\be
\sin \theta = \gamma,
\label{A3}
\ee
and $\Omega$ is an arbitrary real parameter which will be conveniently chosen later.
The transformation \eqref{A40} casts 
equations \eqref{A1} in the form
\begin{align}
i U_t+ U_{xx}-\Omega U + 2|U|^2 U= - \cos{\theta} \,  V + i \gamma (U-V), \nonumber
\\
iV_t+ V_{xx} -\Omega V + 2 |V|^2V= -\cos{\theta} \, U+  i \gamma (U- V).
\label{A6}
\end{align}

The system \eqref{A6} admits an obvious reduction  $U=V \equiv \phi$ 
to the scalar cubic  Schr\"odinger equation,
\be
i \phi_t +\phi_{xx} -a^2 \phi  + 2|\phi|^2 \phi=0,
\label{A3p}
\ee
where $a^2=\Omega -\cos \theta$. 
The relation \eqref{A3} defines two different angles,
$\theta=\arcsin \gamma$ and $\theta= \pi-\arcsin \gamma$.
Accordingly,  Eq.\eqref{A3p} describes 
 two separate invariant manifolds of 
 the system \eqref{A1}.
 Both invariant manifolds are characterised by the Hamiltonian evolution.

Equation \eqref{A3p} has   a family of stationary soliton solutions. Without loss of generality we can restrict 
ourselves to {\it time-independent\/} solutions, 
\[
\phi(x)=    a \sech (ax).
\] 
These define two coexisting families of stationary solitons of the original system \eqref{A1},
with arbitrary amplitudes
  $a>0$  and the corresponding frequencies
\[
      \Omega =   a^2   + \cos \theta.
 \]

The two families of solitons are
distinguished by the sign of $\cos \theta$. 
  One has
$\cos \theta =\sqrt{1-\gamma^2}>0$; we will be denoting the corresponding solitons by ${\vec \psi}_+ = (u_+, v_+)$.
The other one [denoted ${\vec \psi}_- =(u_-, v_-)$ in what follows] has 
$\cos \theta =-\sqrt{1-\gamma^2}<0$. Note that  for the given amplitude $a$, the frequency $\Omega$ corresponding
to the soliton ${\vec \psi}_+$ is higher than that for ${\vec \psi}_-$.
For this reason, we will be referring to the two solitons as 
`high frequency' and `low frequency' solitons. (It is fitting to note that the previous authors  \cite{Suchkov:2011-46609:PRE} considered 
the high-frequency soliton only.)

We note that either family exists only if 
\be
\gamma<1.
\label{A5}
\ee
At the same time, Eq.\eqref{A5} gives the stability condition for the background solution $u=v=0$. In what follows, we will assume that
\eqref{A5} is always imposed.

\section{Stability framework}
\label{Stability}
\subsection{Perturbation decomposition}

To classify the stability of the two families of solitons, we 
 let
\be
U(x,t)= \phi(x)+ \delta U(x,t), \quad 
V(x,t)= \phi(x)+  \delta V(x,t)
\label{K30}
\ee
and  linearise Eqs.\eqref{A6} in $\delta U$ and $\delta V$.
As we will see, a special role is played by the symmetric and 
antisymmetric combinations
\be
p=\frac{\delta U+ \delta V}{\sqrt2}, \quad q=\frac{ \delta U-\delta V}{\sqrt2}.
\label{B6}
\ee

Since  the linearised equations are autonomous in  time,
it is sufficient to consider separable solutions of the form
\begin{align}
p= e^{\nu t} [(p_1' +ip_2') \cos \omega t + (p_1'' + ip_2'') \sin \omega t]  \nonumber  \\
q= e^{\nu t} [(q_1' +iq_2') \cos \omega t + (q_1'' + iq_2'') \sin \omega t],   \label{B5}
\end{align} 
where we have introduced real components of four complex functions $p_1(x)$, $p_2(x)$, $q_1(x)$
and $q_2(x)$:
\begin{align*}
p_1= p_1'+ip_1'', \quad p_2= p_2'+ip_2'', \\
q_1=q_1'+ iq_1'', \quad q_2=q_2'+ iq_2''.
\end{align*}
In \eqref{B5}, both $\nu$ and $\omega$ are assumed to be real. 

Substituting \eqref{B5} in the linearised equations
yields an eigenvalue problem 
\begin{subequations} \label{B7}
\begin{align}
(\mathcal{L} - \cos \theta) {\vec p}  + 2 \gamma J {\vec q}  = \mu J {\vec p},  \label{A7} \\
(\mathcal{L}+ \cos \theta) {\vec q} = \mu J {\vec q}  \label{A8}
\end{align} \end{subequations} 
for two-component complex vectors 
\[
{\vec {p}}= \left( \begin{array}{c} p_1 \\ p_2 \end{array} \right), \quad
{\vec {q}}= \left( \begin{array}{c} q_1 \\ q_2 \end{array} \right).
\]
In \eqref{A7}-\eqref{A8} we have defined $\mu=\nu-i \omega$
and introduced the operator
\[
\mathcal{L}= \left( \begin{array}{cc}
-d^2/dx^2+ \Omega -6\phi^2 & 0 \\ 0 & -d^2/dx^2+ \Omega -2\phi^2
\end{array} \right).
\]
The $J$ in the right-hand sides of \eqref{A7}-\eqref{A8} stands for a skew-symmetric matrix:
\[
J= \left( \begin{array}{lr} 0 & -1 \\
1 & 0 \end{array} \right).
\]

The rotation \eqref{B6} did not diagonalise the $2 \times 2$  block
supermatrix in the left-hand side of \eqref{B7};
however it brought it to the triangular form.
The triangular block matrix 
\be
\left( 
\begin{array}{cc}
\mathcal{L} - \cos \theta  & 2 \gamma J \\
0 &  \mathcal{L} + \cos \theta
\end{array}
\right)
\label{B4}
\ee
has two eigenvectors. One is 
\be
\left( \begin{array}{c} {\vec p} \\ 0  \end{array} \right),
\label{K31}
\ee
 with ${\vec p}$ satisfying
\be
( \mathcal{L} - \cos \theta) {\vec p}= \mu J {\vec p}.  \label{A9}
\ee
This is nothing but the linearised eigenvalue problem for the unperturbed cubic Schr\"odinger
equation (the integrable nonlinear Schr\"odinger equation). 
(Thus ${\vec p}$ is the component of the perturbation that lies in the tangent space to the 
conservative manifold containing the soliton.) The 
 spectrum of $\mu$ consists of a four-fold zero eigenvalue and 
 the continuous spectrum which lies on the imaginary axis;
 there are no unstable $\mu$'s here.

The second eigenvector, 
\be
\left( \begin{array}{c} {\vec p} \\ {\vec q} \end{array} \right),
\label{K32} 
\ee
 includes a nonzero ${\vec q}$ component.
These $\vec q$ arise as eigenfunctions of  the operator \eqref{A8}. 
When $\vec q \neq 0$,   
Eq.\eqref{A7} becomes a nonhomogeneous equation with the right-hand side determined by ${\vec q}$:
\be
\frak N {\vec p} =-2 \gamma J {\vec q}.
 \label{A10}
 \ee
 Here $\frak N$ is a nonsymmetric operator defined by 
\[
 \frak N= \mathcal{L} - \cos \theta - \mu J.
 \]
 As discussed in the previous paragraph, the eigenvalue problem \eqref{A9}
does not have nonzero eigenvalues; hence the adjoint operator
\[
\frak N^\dagger = \mathcal{L}  - \cos \theta + \mu J
\]
  has null eigenvectors  only if $\mu=0$. 
Therefore, for any $\mu \neq 0$, Eq.\eqref{A10} has a bounded solution. 

Thus the stability analysis reduces to solving the eigenvalue problem \eqref{A8}. 
It is important to emphasise that, despite the presence of 
gain and loss, the eigenvalue problems \eqref{A8} and \eqref{A9} are symplectic
(that is, pertaining to Hamiltonian evolutions). 
The implication is that  stable limit cycles
are not the objects that can be expected to bifurcate from stationary solitons when the
latter lose their stability. That is, the instability should evolve according to scenarios that are 
characteristic for conservative systems, e.g. breakup into long-lived transient structures,
singular growth etc.

It  is also worth commenting on the significance of the triangular 
representation \eqref{B7} which decomposes 
 small perturbations  into the part tangent to the 
conservative invariant manifold containing the soliton, and the part 
that is transversal to this manifold. This decomposition alone
explains some numerical observations of the previous authors \cite{Suchkov:2011-46609:PRE},
in particular the instability of the bond-centred solitons in the discrete case. 
Indeed, the instability of the bond-centred soliton solutions of  the discretised Eq.\eqref{A1} is
caused by perturbations lying in the conservative invariant manifold
(that is, satisfying the scalar nonlinear Schr\"odinger equation). 
The instability of the bond-centred vector solitons is  simply inherited from the instability
of their scalar counterparts.

\subsection{Integral considerations}

It is instructive to  consider the effect of the perturbations on the Hamiltonian, momenta and the power integrals.
Substituting  Eqs.\eqref{K30}-\eqref{B5}  in \eqref{PP}, \eqref{MM} and \eqref{FF},  the right-hand sides of
\eqref{P}, \eqref{M} and \eqref{H} are evaluated to be
\begin{subequations}
\label{delta} 
\begin{align}
\left. (\mathcal{P}_u-\mathcal{P}_v) \right|_{t=0} = 2 \sqrt{2} \int q_1'(x)                 \phi (x)           dx,  \label{PPP} \\
\left. ({\mathcal M}_u-{\mathcal M}_v) \right|_{t=0} = -2 \sqrt{2} \int q_2'(x)               \phi_x             dx, \\
\left. ({\mathcal R}_u-{\mathcal R}_v) \right|_{t=0}  = -2 \sqrt{2} \int   \! q_1'(x)  (a^2+ 2 \phi^2)  \phi \, dx.
 \end{align}
 \end{subequations}
In Eqs.\eqref{delta}   we kept terms only up to the linear order in $p_{1,2}$ and $q_{1,2}$.

The eigenvector \eqref{K31}   has $q_1'=q_2'=0$; the corresponding rates \eqref{delta} are all zero. 
The perturbations associated with eigenvectors of this type do not trigger the growth or decay of 
 the total power, momentum and the Hamiltonian. They just take the soliton to a nearby solution of the 
 scalar Schr\"odinger equation \eqref{A3p};  the $\mathcal{PT}$ symmetry 
 (the symmetry between the $u$ and $v$ components) remains unbroken.

As for the eigenvectors of the second type, Eqs.\eqref{K32} with nonzero ${\vec q}$ and $\vec p$ satisfying
\eqref{A8} and \eqref{A10},  they may set  nonvanishing rates of change of the total power, momentum and Hamiltonian.
 Whether the right-hand side of Eq.\eqref{P}, \eqref{M} or  \eqref{H}  is zero or not, depends, in particular,  on the parity of
 the eigenfunction ${\vec q}(x)$ of the symplectic operator in \eqref{A8}. 
 If $({\vec p}, {\vec q})^T$ is an eigenvector associated with an unstable eigenvalue ($\mathrm{Re} \, \mu>0$)
 and such that a particular
 right-hand side in Eq.\eqref{P}, \eqref{M} or  \eqref{H} is nonzero, 
 the corresponding integral in the left-hand side  
  will start evolving away from its soliton value.
  
In what follows, the  perturbations  of the integrals pertaining to            the        individual            $u$ and $v$ components
will also prove useful.
% The perturbations of the momenta will prove particularly useful in what follows.
Restricting ourselves to perturbations associated with real eigenvalues $\mu>0$,
and substituting Eqs.\eqref{K30}-\eqref{B5} in  \eqref{P} and \eqref{M}, we arrive at
\begin{subequations} \label{P4}
  \begin{align}
  \delta {\mathcal P}_u= \sqrt{2}  \left( \frac{ 2 \gamma}{\mu}+1 \right)  e^{\mu t} \int q_1' \phi \,  dx,   \label{P4a} \\
 \delta {\mathcal P}_v= \sqrt{2}  \left( \frac{ 2 \gamma}{\mu}- 1 \right)  e^{\mu t} \int q_1' \phi \, dx, \label{P4b}
 \end{align}
 \end{subequations}
 and
 \begin{subequations} \label{M4}
 \begin{align}
   {\mathcal M}_u= - \sqrt{2} \left( \frac{ 2 \gamma}{\mu}+1 \right)  e^{\mu t} \int q_2' \phi_x dx,    \label{M4a}  \\
 {\mathcal M}_v = - \sqrt{2} \left( \frac{ 2 \gamma}{\mu} -1 \right)  e^{\mu t}  \int q_2' \phi_x dx. \label{M4b}
\end{align}
\end{subequations}

\subsection{Symplectic eigenvalue problem}
\label{symp}

Defining $X= ax$, $\lambda= \mu/a^2$, and introducing 
\be
\eta= 2 \frac{\cos \theta}{a^2},  \label{E}
\ee
the problem \eqref{A8} can be written as
\be
\left(\begin{array}{cc}
L_1+ \eta & 0 \\ 0 & L_0+ \eta
\end{array} \right)
\left( \begin{array}{c} g \\ f \end{array} \right) = \lambda J \left( \begin{array}{c} g \\ f \end{array} \right).
\label{A11} 
\ee
Here $L_{0,1}$ stand for the scalar Sturm-Liouville operators
\begin{align}
L_0= -d^2/dX^2+ 1 - 2 \sech^2X, \nonumber 
\\
L_1= -d^2/dX^2+ 1 - 6 \sech^2X, \label{B12}
\end{align}
and we have redenoted $q_1=g$ and $q_2=f$ for notational convenience.
The eigenvalues $\lambda$ and eigenvectors $(g,f)$ are generally complex.
Positive $\eta$ correspond to the soliton ${\vec \psi}_+$ and negative to ${\vec \psi}_-$.

Thus we have reduced a two-parameter stability problem 
to an eigenvalue problem involving
a single similarity parameter. Solitons with different amplitudes and in systems with different
gain-loss coefficients have the same stability properties as long as they share the value of $\eta$. 
(It is fitting to note that a self-similarity of this sort was previously encountered in the parametrically
driven damped nonlinear Schr\"odinger equation \cite{Barashenkov:1991-113:EPL,Barashenkov:2002-104101:PRL}. The difference of the parametrically driven
situation from the present setting was that there, the similarity combination included two control parameters of the
equation whereas Eq.\eqref{E} combines a control parameter with a free amplitude of the soliton.)

The scalar operators have familiar spectral properties. The only discrete eigenvalue of $L_0$ is zero; the 
associated eigenfunction is even: $L_0 z_0=0$, $z_0= \sech X$. The operator does not have other eigenvalues between 
0 and 1 (the edge of the continuous spectrum).  The lowest eigenvalue of 
$L_1$ is $-3$; the associated eigenfunction is $y_0=\sech^2 X$. The only other eigenvalue is 0;
the corresponding eigenfunction $y_1= \sech X \tanh X$ is odd.

 \begin{figure}[t]
 \begin{center} 
\includegraphics*[width=\linewidth]{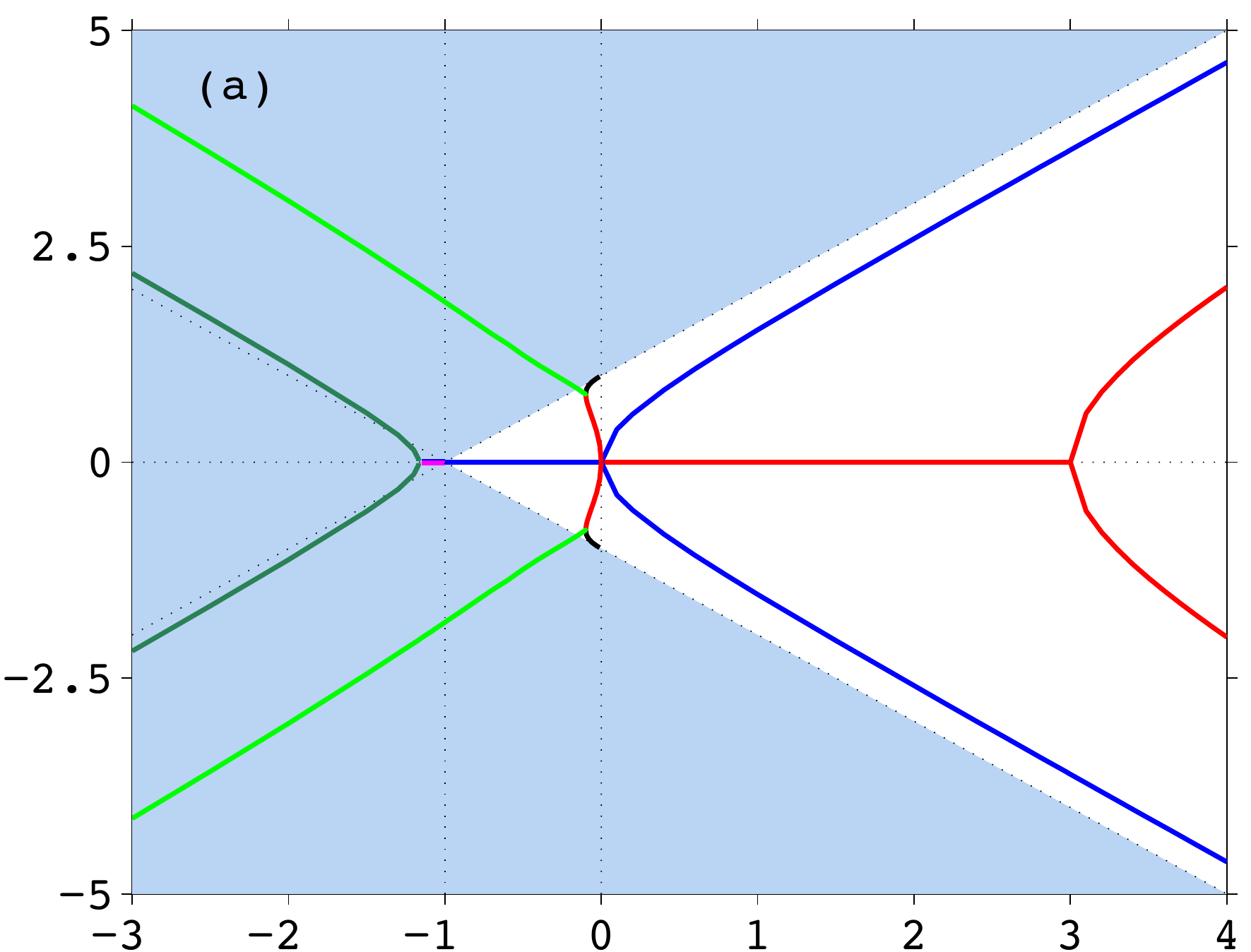} % {darkblue.eps}
 \includegraphics*[width=\linewidth]{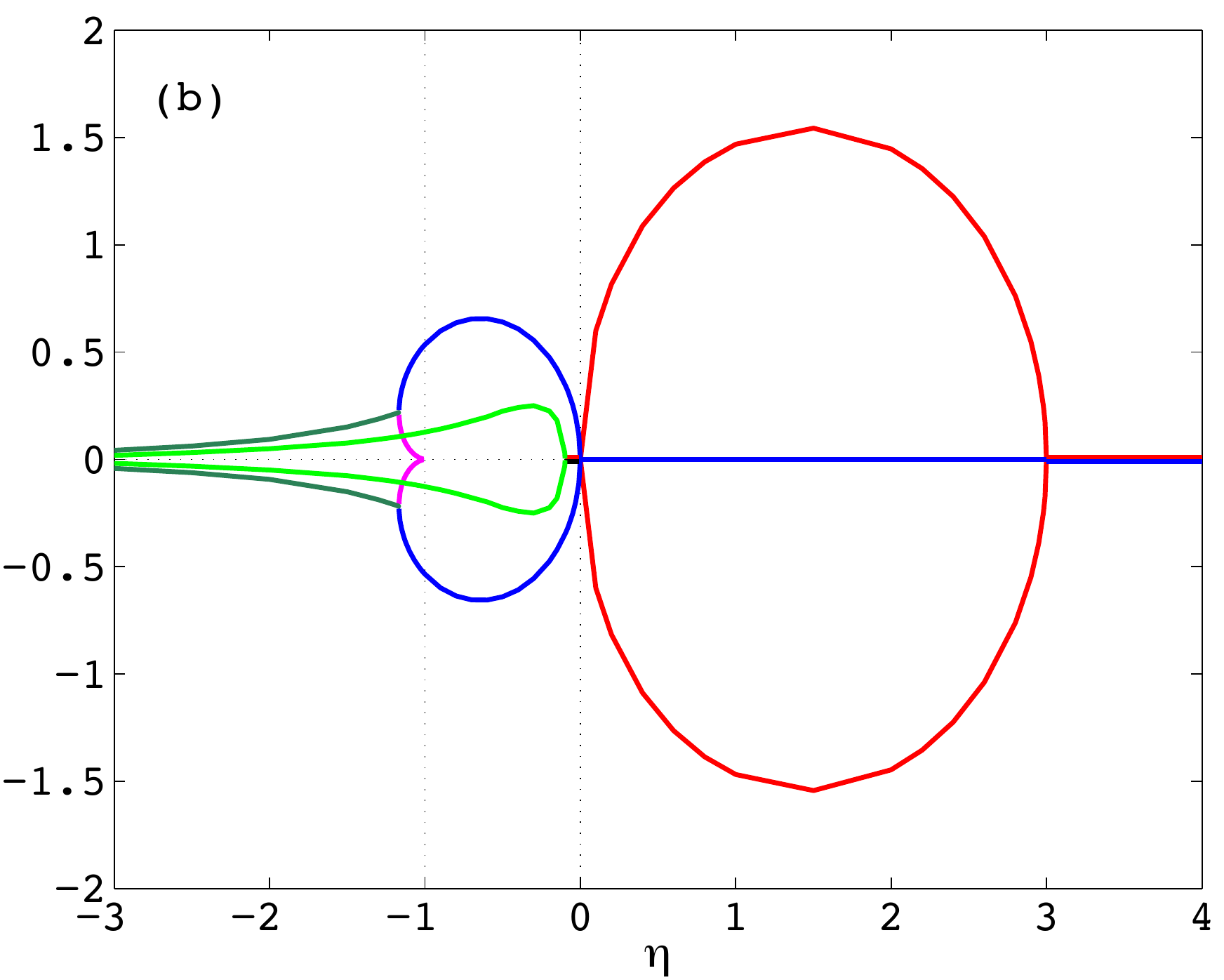} % {fig1b.eps} % {real11.eps}
  \caption{(a) the imaginary and (b) real  part of the eigenvalue $\lambda$ of the
   problem \eqref{A11} as a function of ${\eta}$.
   (For the fixed $\gamma$, the parameter $\eta$ is proportional to the inverse amplitude squared: $\eta =2 \cos \theta/a^2$.)
    A real  pair of eigenvalues moves onto the imaginary axis while
   the imaginary pair becomes real as ${\eta}$ passes through zero
   (either way). In the ${\eta}>0$ part of the figure, the real pair subsequently converges and returns
   to the imaginary axis, so that for ${\eta} \geq 3$, all eigenvalues are pure imaginary. In the ${\eta}<0$ part, two complex
   quadruplets are born in two consecutive Hamiltonian Hopf bifurcations. The imaginary parts of the quadruplets grow as ${\eta} \to -\infty$
   while the real parts decay. In (a), tinted are the regions $\mathrm{Im} \, \lambda \geq \eta+1$ and $\mathrm{Im} \, \lambda \leq -(\eta+1)$ 
   filled with the continuous spectrum.}
 \label{EV_cinfty}
 \end{center}
 \end{figure}

When $\eta=0$, the eigenvalue problem \eqref{A11} coincides with the eigenvalue problem
for the unperturbed cubic nonlinear  Schr\"odinger equation. It is important to emphasise, however, that  the choice ${\eta}=0$ 
does not correspond to the undamped-undriven situation. % (where $\gamma=0$).
(The undamped-undriven limit $\gamma=0$ does not single out  any particular ${\eta}$ and is not special in any way.)
What the value ${\eta}=0$  corresponds to, is the turning point  $\gamma =  1$.  As $\gamma$ 
reaches 1 from below, the solitons ${\vec \psi}_+$ and ${\vec \psi}_-$  merge and disappear.

As in the unperturbed nonlinear Schr\"odinger, there is a four-fold zero eigenvalue at the point ${\eta}=0$:
$\lambda^{(1,2,3,4)}=0$.
As ${\eta}$ deviates from zero,  the four eigenvalues move out of the origin.
A pertubation calculation (Appendix \ref{RIT})
shows that two opposite eigenvalues move on to
the imaginary axis, and the other two move on to the positive and negative
real axis, respectively:
\be
\lambda^{(1,2)} = \pm 2 {\eta}^{1/2} + O({\eta}), \quad
\lambda^{(3,4)}= \pm \frac{2}{\sqrt{3}} (-\eta)^{1/2} + O(\eta).
\label{X0}
\ee

The real and imaginary parts of  eigenvalues of the problem \eqref{A11}, obtained numerically,
 are plotted in fig.\ref{EV_cinfty}.
 The two pairs of real and imaginary eigenvalues emerging from the origin, are clearly visible.

Before proceeding to the evolution of the eigenvalues as ${\eta}$ grows to large positive respectively
negative values, 
it is appropriate to note the position of the continuous spectrum
the symplectic operator 
\be
J^{-1} \left(
\begin{array}{cc}
L_1+{\eta} & 0  \\ 0 & L_0+{\eta} \end{array}
\right)
\label{Z0}
\ee
in \eqref{A11}. 
There are two branches of 
continuous spectrum, both lying on the imaginary axis of $\lambda$:  $\lambda= \pm i \omega(k)$,
where $\omega= 1+{\eta}+k^2$. 
(These are indicated by shading in fig.\ref{EV_cinfty}(a).)
When ${\eta} >-1$, the continuous spectrum has a gap, $[-i(1+{\eta}), i(1+{\eta})]$.

It is also worth mentioning that the  spectrum of symplectic operators 
consists of pairs of opposite pure-imaginary values, real pairs and
complex  quadruplets. If $\lambda$ is a real or pure imaginary point
of the spectrum, then $-\lambda$ is another one;  if a complex $\lambda$ is in the spectrum, 
then so are $-\lambda, \lambda^*$, and $-\lambda^*$ \cite{Arnold:2010:MathematicalMethods}.

\section{High frequency soliton}
\label{High}

 \begin{figure}[t]
 \begin{center} 
  \includegraphics*[width=\linewidth]{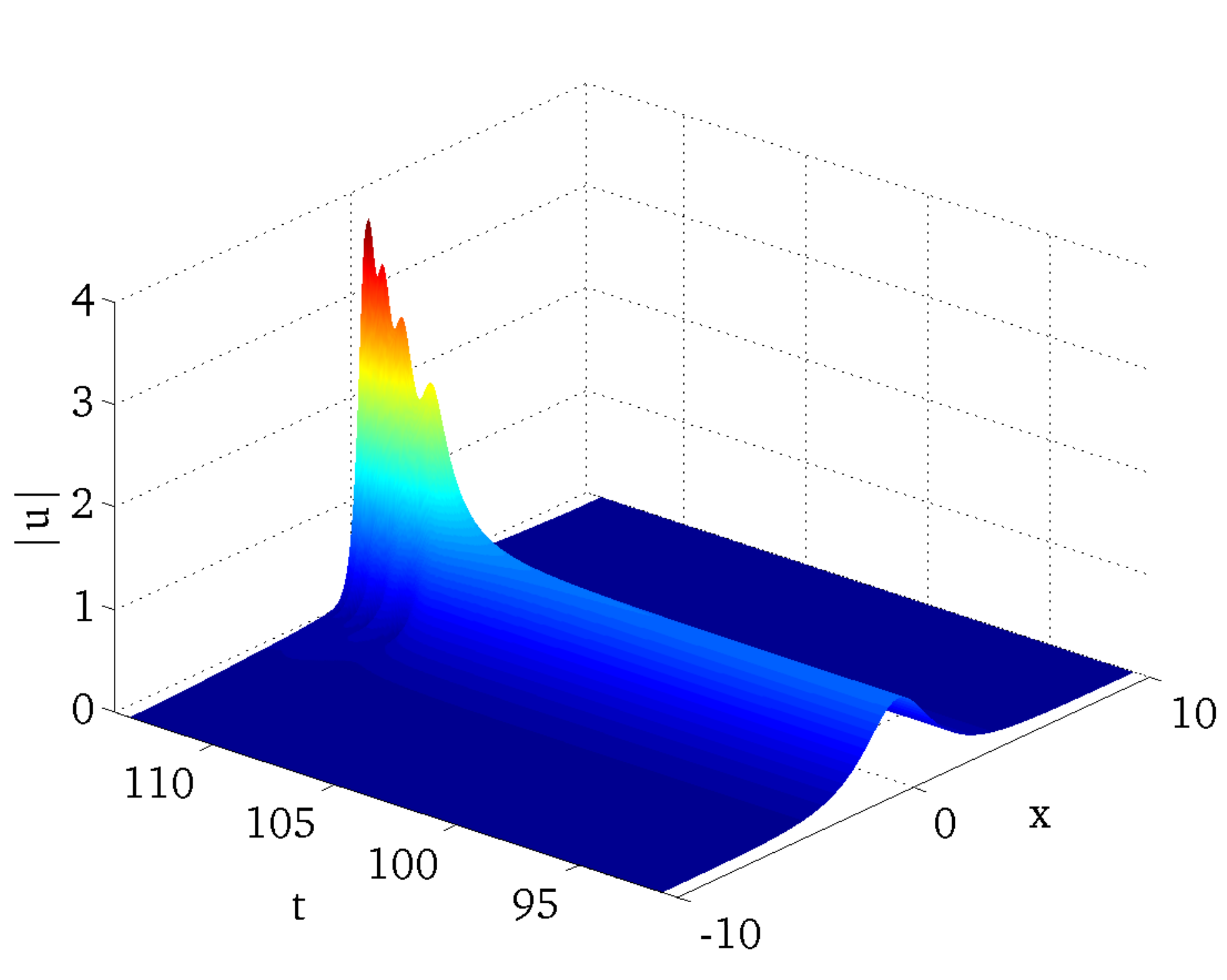} % {fig2a_short.eps}
 \includegraphics*[width=\linewidth]{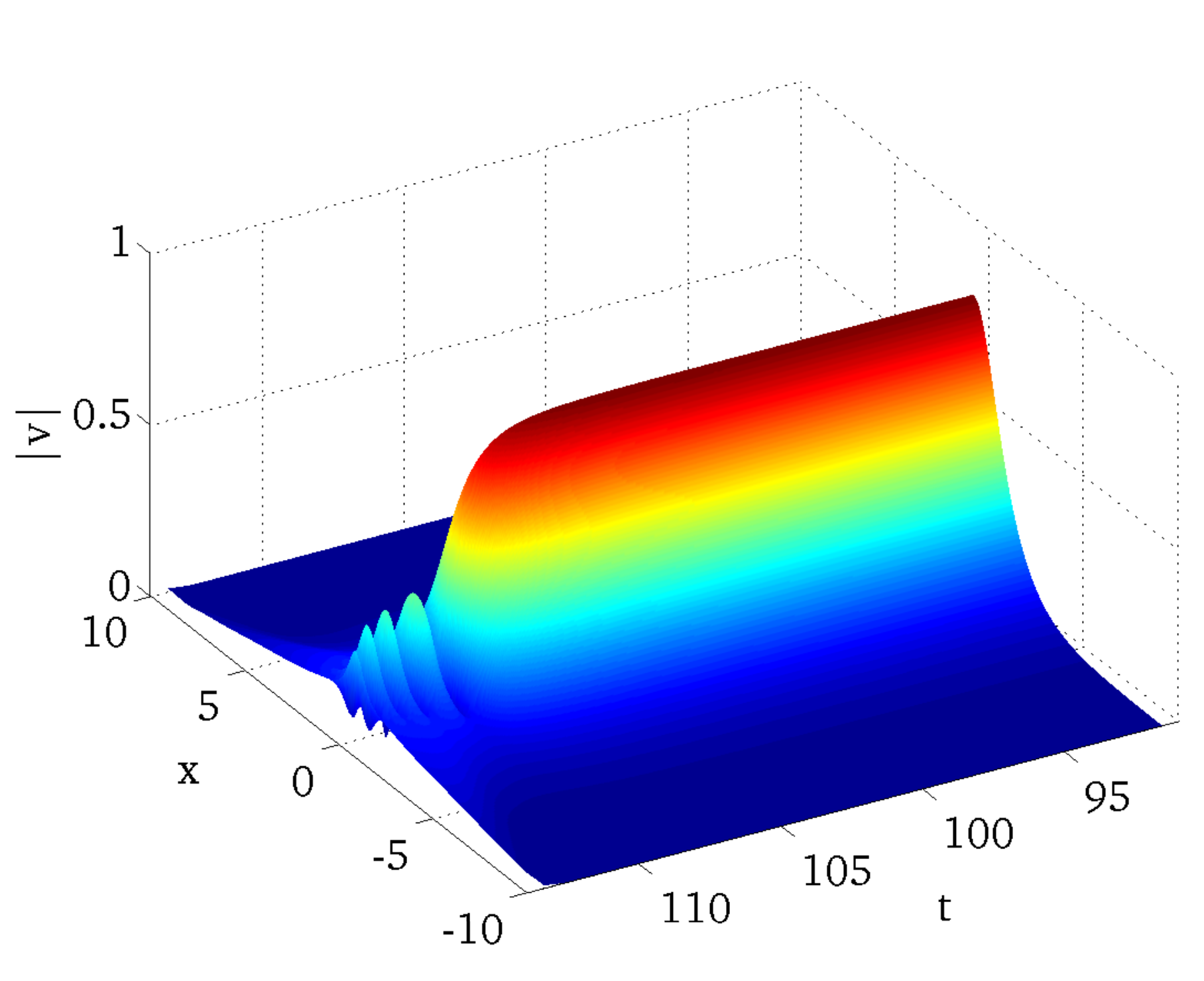} % {fig2b_short.eps}   
 \caption{The evolution of  the perturbed soliton ${\vec \psi}_+$ with the amplitude $a$ above the threshold $a_c$. 
  Shown is the magnitude of the field $u$ (a) and field $v$ (b). 
  The control parameter is set to $\gamma=0.1$ (for which $a_c=0.814$); the soliton's amplitude is $a=0.820$.
 Note that the spatial interval has been cut down to $(-10,10)$ for visual clarity
 and that only the late stage of evolution is shown.
 In both panels the colour varies from deep blue
  (lowest elevation)  to deep red (highest elevation);  since
  the maximum values of $|u|$ and $|v|$ are not equal, the same colour corresponds to different values in the two panels. 
  The same convention is used in all other double-panel 3D plots.
   \label{ev_Plus}}
 \end{center}
 \end{figure}

 \begin{figure}[t]
 \begin{center} 
   \includegraphics*[width=\linewidth]{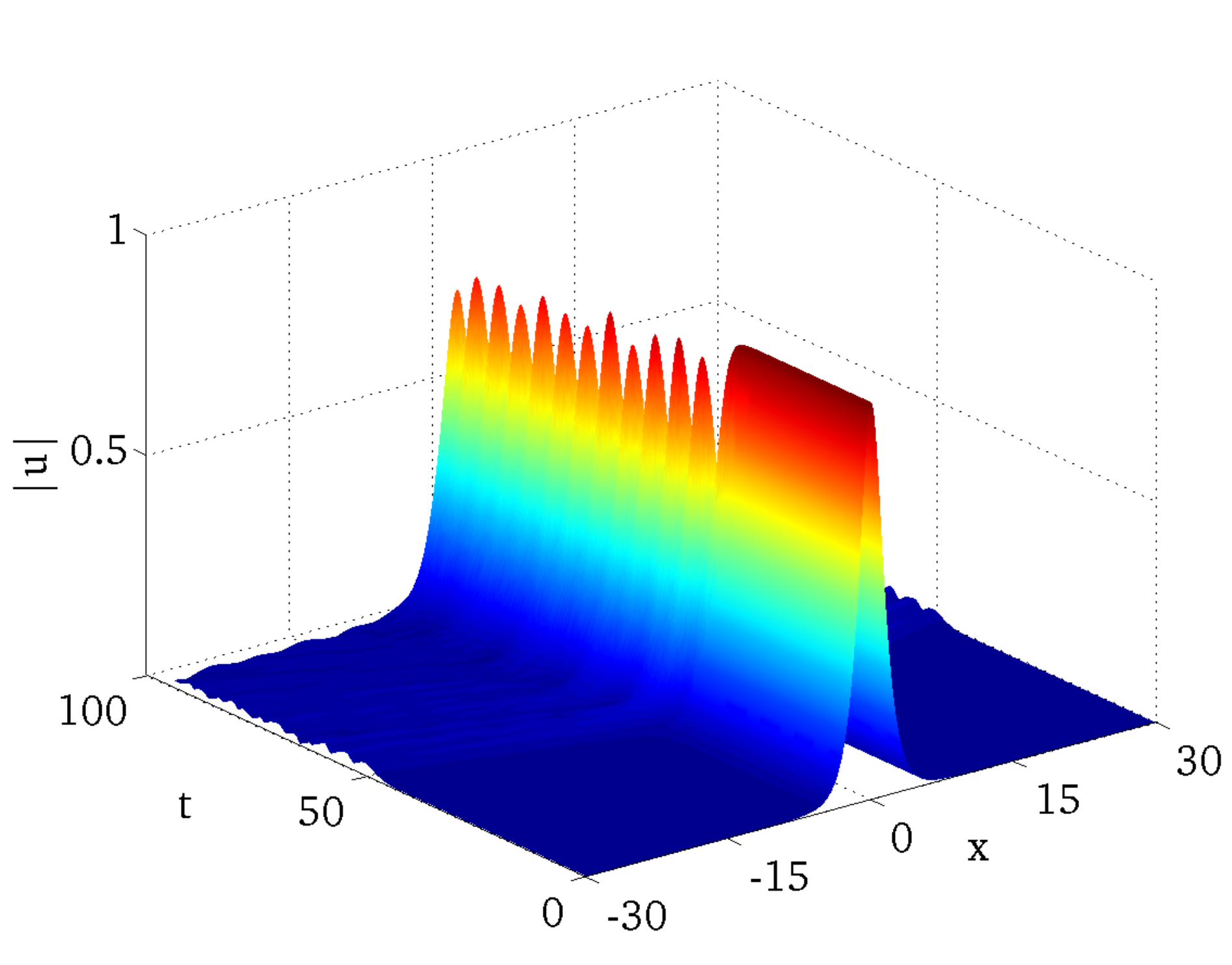} % {psiplus_fig4_u.eps}
 \includegraphics*[width=\linewidth]{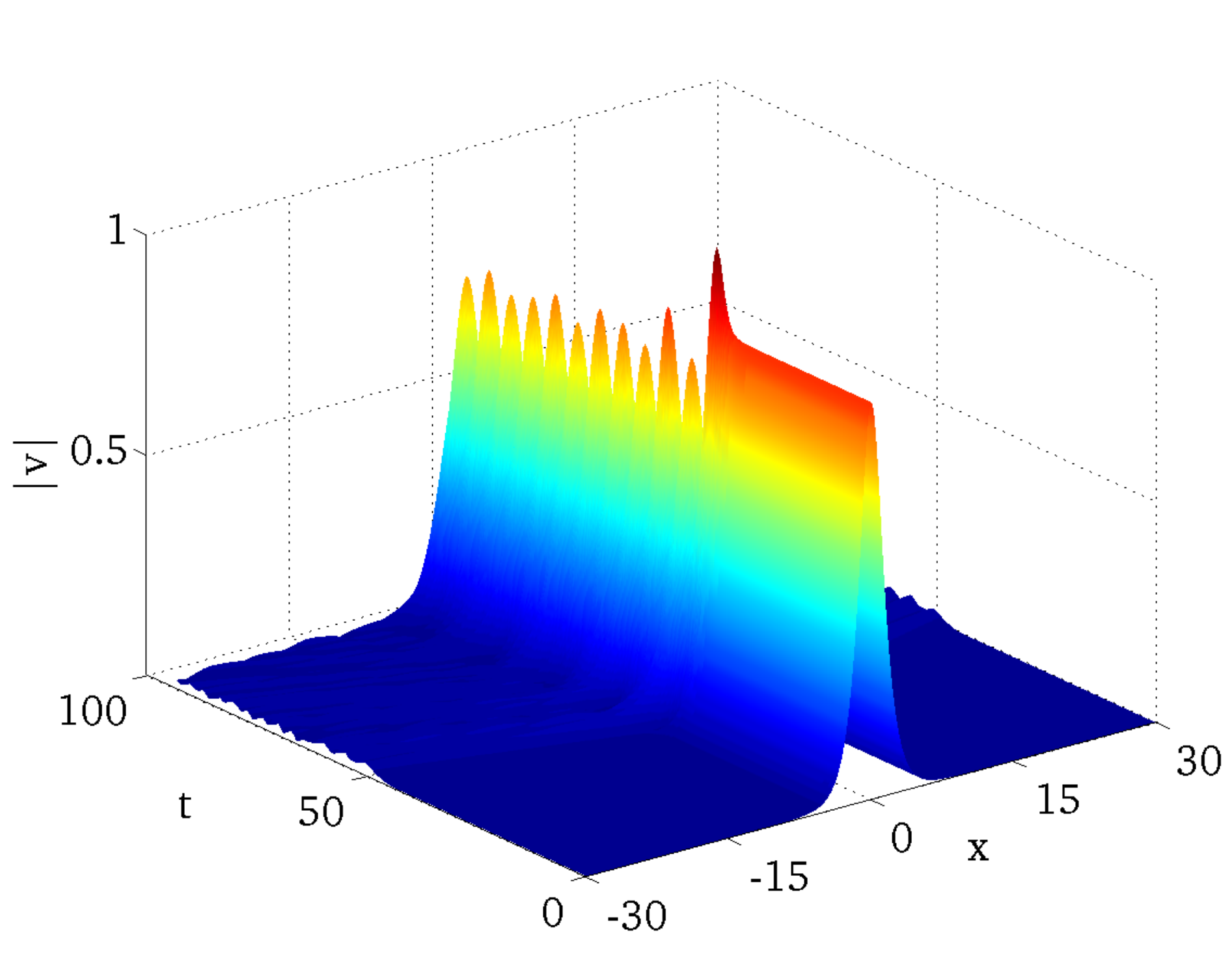} % {psiplus_fig4_v.eps}
    \caption{The evolution of  the unstable soliton ${\vec \psi}_+$ may result in the formation of a breather.
  Shown is the magnitude of the field $u$ (a) and field $v$ (b). 
  Here $\gamma=0.1$ and $a=0.9$.
   \label{ev_Plus_g01_a09}}
 \end{center}
 \end{figure}

In the case of the soliton ${\vec \psi}_+$, the eigenvalue problem is amenable to simple analysis.
The lowest eigenvalue of the operator $L_0+\eta$ equals ${\eta}$; therefore, when ${\eta}>0$,
the operator $L_0+{\eta}$ is positive definite and admits an inverse.
The problem \eqref{A11} can be written then as a generalised
eigenvalue problem for a scalar function $g(X)$:
\be
(L_1+{\eta})g= -\lambda^2 (L_0+{\eta})^{-1} g.
\label{A12}
\ee

The operator on the left in \eqref{A12} is symmetric, and the one on the right
is symmetric and positive definite. The  lowest eigenvalue
$-\lambda^2$ in \eqref{A12} is given by the minimum of the Rayleigh quotient:
\be
-\lambda^2 = \min \frac{\langle g |L_1+{\eta}|g \rangle}{\langle g |(L_0+{\eta})^{-1}  |g \rangle}.
\label{A13}
\ee
[Here the bra-ket notation is used for the $\mathcal{L}^2$ scalar product:
 $\langle y | z \rangle = \int y(X) z(X) dX$.]
The minimum is positive if the lowest eigenvalue of the 
operator  in the numerator ($\nu=-3+{\eta}$)  is positive: ${\eta}>3$. 
Recalling the definition \eqref{E}, we arrive at the stability
condition for the soliton ${\vec \psi}_+$:
\be
a \leq a_c,  \quad a_c^2 = \frac23 \sqrt{1-\gamma^2}.
\label{A14}
\ee

  \begin{figure}[t]
 \begin{center} 
\includegraphics*[width=\linewidth]{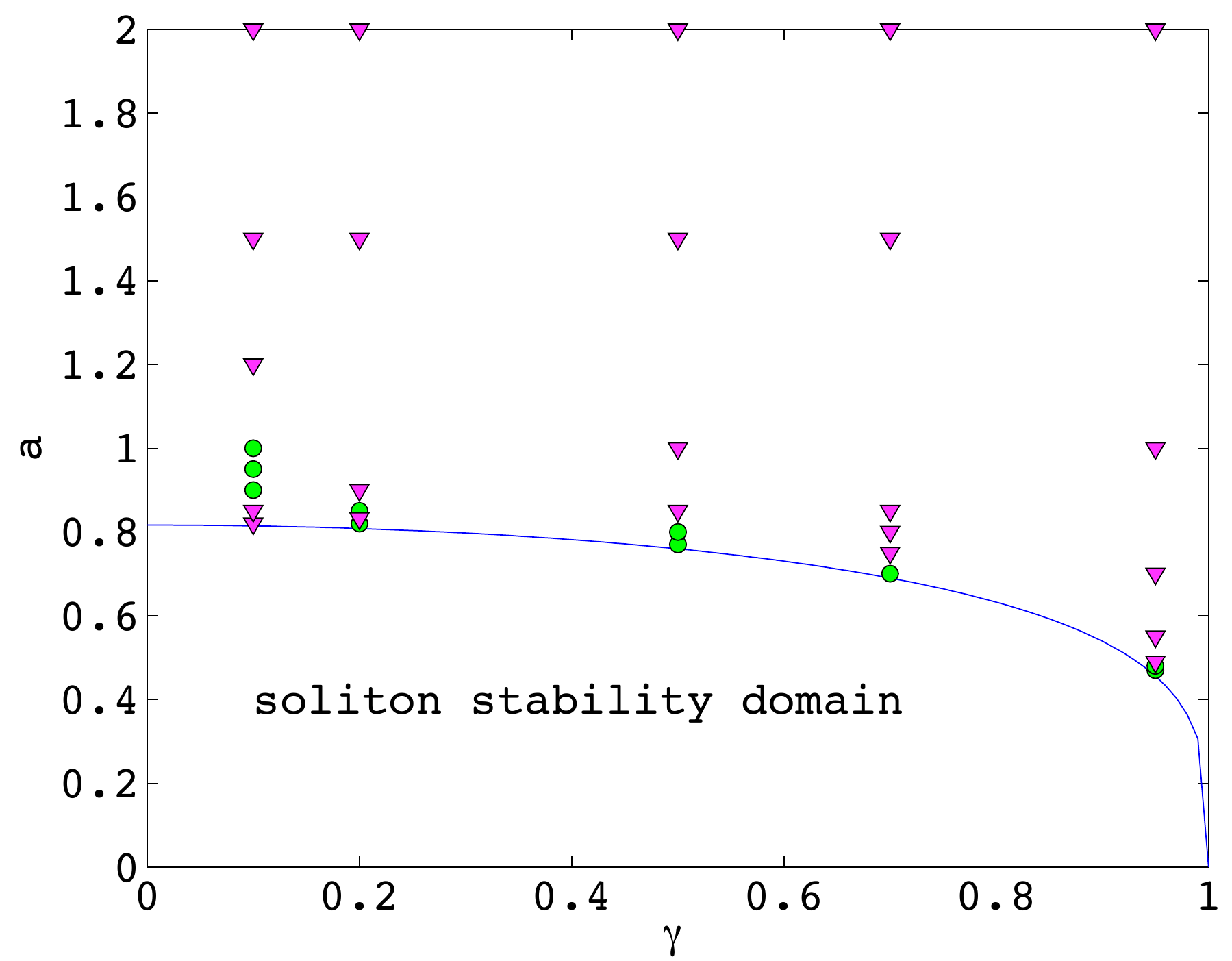} %{new_psiplus_chart.eps}
    \caption{Chart of asymptotic regimes emerging from the unstable soliton ${\vec \psi}_+$.
  The five columns correspond to $\gamma=0.1, 0.2, 0.5, 0.7$ and $0.95$.
  The solid line demarcates the boundary \eqref{A14}; the ${\vec \psi}_+$ soliton is unstable above this line.
  The magenta triangles mark blowups while green circles indicate the formation of breathers.
 \label{chart_plus}}
 \end{center}
 \end{figure}

The numerical study of the eigenvalue problem \eqref{A11} corroborates these conclusions. As 
${\eta}$ grows from 0 to positive values, two pairs of opposite eigenvalues, a real and a pure imaginary pair,  appear from the origin 
on the $(\mathrm{Re} \, \lambda, 
\mathrm{Im} \, \lambda)$ plane.
The real eigenvalues are associated with even and imaginary pair with odd eigenfunctions.
  The two pure imaginary eigenvalues diverge to their respective infinities (fig.\ref{EV_cinfty}(a)).
The real pair first grows in absolute value, but then the real eigenvalues
reverse (fig.\ref{EV_cinfty}(b)) and, as ${\eta}$ reaches $3$,  collide at the origin and move on to the imaginary axis.
The emerging second pair of imaginary eigenvalues diverges to the infinities, like the first pair before (fig.\ref{EV_cinfty}(a)).

The quantity $q(X,t)$ in \eqref{B6} measures the difference between the two components of the vector $(U,V)$.
If the eigenvalue problem \eqref{A11} has a positive
eigenvalue $\lambda$, the difference grows monotonically:
$q=(q_1+iq_2)e^{\lambda a^2 t}$. Therefore, the instability associated with real eigenvalues
should manifest itself as a monotonically
growing asymmetry between the two components of the vector field.
Quantitatively, this asymmetry is measured by the difference in the increments
of the ${\mathcal P}_u$ and ${\mathcal P}_v$ integrals,  Eqs.\eqref{P4}.

Transient and asymptotic solutions emerging as the perturbation grows, should inherit
the spatial parity of the eigenvector $(g(X), f(X))$ associated with the real eigenvalue
--- that is, should be even in $X$.

The imaginary eigenvalues correspond to internal modes of the soliton. In  Appendix \ref{asympt}, we 
derive the asymptotes for both pairs of imaginary eigenvalues as ${\eta} \to \infty$:
$\lambda = \pm i (\eta+ 0.685)+ O(1/\eta)$ and $\lambda = \pm i (\eta-~1.438)+ O(1/\eta)$.

Figs \ref{ev_Plus} and \ref{ev_Plus_g01_a09} 
present results of direct numerical simulations of
 the ${\vec \psi}_+$ soliton. For five values of the control parameter, 
$\gamma=0.1$, $0.2$, $0.5$, $0.7$, and $0.95$, we simulated solitons with amplitudes $a$ above the 
critical one, $a_c$, given by \eqref{A14}. 
 In each of the five cases we have identified two possible scenarios of instability growth.
 In one of these, 
 the magnitude of the field $u$ grows  without bound,  while $v$ decreases.
(See fig.\ref{ev_Plus}.)  We will be referring to this type of evolution as the `blowup'. 
Note that this asymmetry-growth scenario is in agreement with our 
expectations based on the eigenfunction analysis.

   In the other scenario,
the breakup of the unstable soliton  results 
 in the formation of a long-lived oscillatory state --- a kind of a breather (fig. \ref{ev_Plus_g01_a09}).
Here, the initial stage of the evolution is also characterised by the growth of asymmetry.
Unlike  fig.\ref{ev_Plus}, it is the $v$ component that is growing this time, and $u$ is the one that is decreasing
(clearly visible in fig. \ref{ev_Plus_g01_a09}). This ``anomalous" growth cannot continue indefinitely
[this would contradict 
Eq.\eqref{P}] and eventually the evolution is captured into the breather regime.

 Fig.\ref{chart_plus} sketches the  ranges of $a$ which are characterised by each of these two
 types of instability growth. Typically, solitons with amplitudes only slightly exceeding the threshold 
 \eqref{A14} give rise to breathers whereas 
 the blow-up scenario is observed for larger $a$. The exception from this rule occurs for small $\gamma \sim 0.1$
 where the blowup and breather domains seem to be interlaced in a more complex fashion.

\section{Low frequency soliton}
\label{Low} 
%\subsection{Numerical eigenvalues for ${\eta}<0$}

When ${\eta}<0$, neither of the scalar operators in \eqref{A11}
is invertible on the $\mathcal{L}^2$ space.  
Here, the analysis  of the 
eigenvalue problem \eqref{A11} has to be done mostly by numerical means.
Our numerical study is summarised in the left half  of fig.\ref{EV_cinfty}.

As ${\eta}$ is decreased through zero, a pair of opposite real eigenvalues 
(associated with even eigenfunctions) collides and moves on to the imaginary axis.
Simultaneously (that is, at ${\eta}=0$)
another pair of opposite imaginary eigenvalues (also with even eigenfunctions)
detaches from the continuous
spectrum. When ${\eta}$ reaches the value $\eta_1$,
\be
\eta_1=-0.0988, 
\label{eta1}
\ee
the four imaginary eigenvalues collide,
pairwise, and leave the imaginary axis forming a complex quadruplet
$\lambda, \lambda^*, -\lambda, -\lambda^*$
(with even eigenfunctions).  As ${\eta}$ grows to larger negative values,
the real parts of the quadruplet first grow but then start decreasing. At the same time, the
imaginary parts grow without bound. 

Another pair of eigenvalues colliding as ${\eta}$ is decreased through zero, is pure imaginary.  Unlike the colliding real pair, the imaginary pair
is associated with odd eigenfunctions. The parity of the eigenfunctions is preserved
as this pair reappears on the real axis for ${\eta}<0$. As
${\eta}$ grows to larger negative values, the absolute values of the  real $\lambda$ first
grow but then start decreasing. 

The existence of the real pair in the interval
$-1<{\eta}<0$ can be established 
analytically. (See  Appendix \ref{E-1}.)
It might be tempting to expect
 the pair to converge at the origin as ${\eta}$ reaches $-1$;
however the actual bifurcation diagram turns out to be more complex
(fig.\ref{EV_cinfty}). In fact
one can {\it prove\/} that $\mathrm{Re} \ \lambda$ remains nonzero at ${\eta}=-1$.
(See Appendix \ref{E-1}.)

In the meantime, as ${\eta}$ is approaching $-1$, the gap in the continuous spectrum 
is shrinking.
As ${\eta}$ reaches $-1$, the two edges of the continuous
spectrum meet at the origin and the gap closes.
Due to the resonance between the edge eigenfunctions, 
 a new pair of discrete eigenvalues is born at this point. 
 Like the coexisting real pair, this pair of opposite real eigenvalues is associated with odd eigenfunctions. 
 As ${\eta}$  drops down  to $\eta_2$, where
 \be
 \eta_2= -1.168, 
 \label{eta2}
 \ee
 the newly born pair of real eigenvalues
 and the real pair that has arrived from the origin collide and emerge into the complex plane. 
 This is where the second complex quadruplet is born.
 As  the negative ${\eta}$ continues to grow in  absolute value, the real parts of the 
 eigenvalues making up the quadruplet decrease,  while the imaginary parts grow.

 Thus when ${\eta}<\eta_2$, we have two complex quadruplets.
 As ${\eta}$ grows in absolute value, the real parts of the eigenvalues in
 both quadruplets decrease whereas 
 imaginary parts grow.
 In the Appendix \ref{asympt}  we derive the
 asymptotic behaviour of the imaginary parts analytically: $\mathrm{Im} \, \lambda = \pm  (\eta-1.438)+ O(1/\eta)$;
 $\mathrm{Im} \, \lambda = \pm (\eta+ 0.685)+ O(1/\eta)$. We also show that the decrease of the
real parts  is {\it exponentially\/} fast. 
Eq.\eqref{E} implies  then that solitons with small amplitude $a$
--- even if are unstable ---
have exponentially long lifetime, 
\[
\tau \sim \frac{1}{a^2}  \exp \left( \frac{\sigma \sqrt{1-\gamma^2} }{a^2} \right),
\]
with some constant $ \sigma>0$.

  \begin{figure}[t]
 \begin{center} 
  \includegraphics*[width=\linewidth]{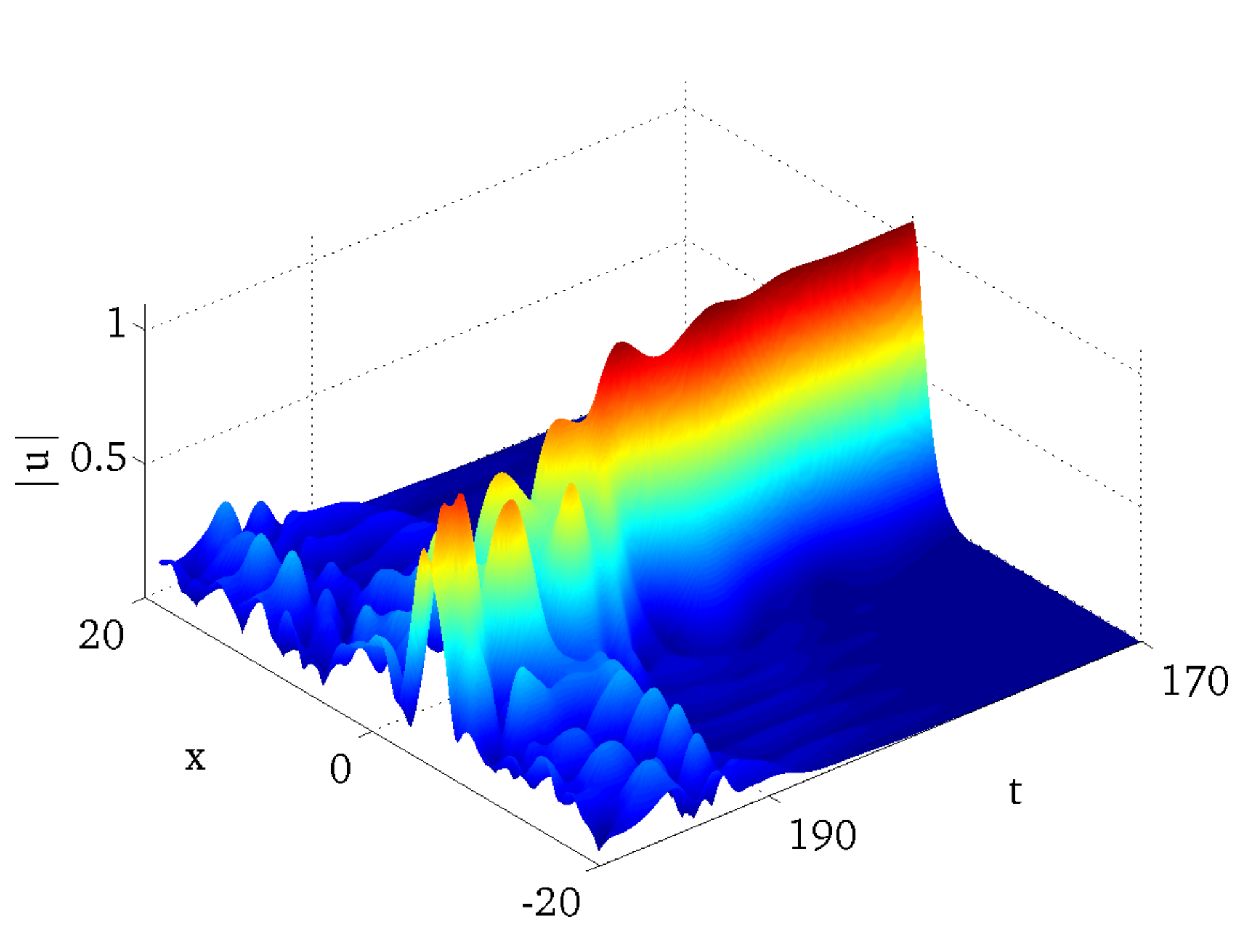} % {psiminus_dissipation_u2.eps}
    \includegraphics*[width=\linewidth]{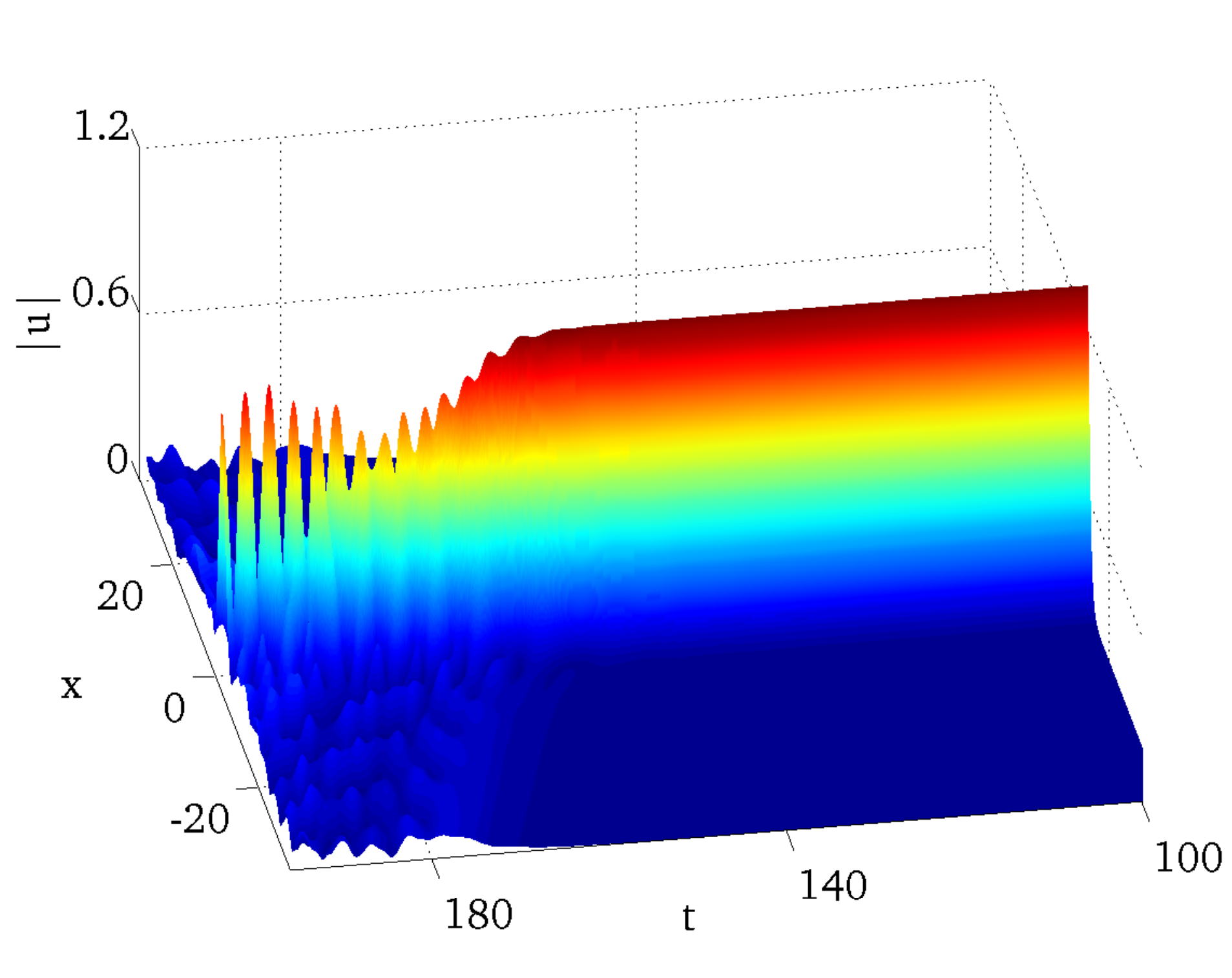} % {psiminus_breather_u2.eps}
  \caption{Unstable $\psi_-$ solitons with $|\eta|>1.168$  disperse (a) or transform into a long-lived breather (b).
  In (a), $\gamma=0.5$ and $a=1.075$;
in (b), $\gamma=0.1$ and $a=1.152$. Both combinations of $a$ and $\gamma$ correspond to  ${\eta}=- 1.5$.   
  Only the $u$-component is shown in both cases; the evolution of the $v$ field is qualitatively similar.
   \label{decay_or_breather}}
 \end{center}
 \end{figure}

The nonlinear evolution of the unstable ${\vec \psi}_-$ soliton is determined by the competition
between the  eigenvalues with positive real parts. 
According to fig.\ref{EV_cinfty}, there are two 
ranges of the soliton amplitude  for each $\gamma$: the small and the large $a$.

For small $a$ [more precisely, for $a$ such that  $|\eta| > |\eta_2|$ with $\eta_2$ as in \eqref{eta2}],
the soliton has two complex quadruplets in its spectrum ---
one  with an even and the other one with an odd eigenvector.
Accordingly,  the instability growth of a small-amplitude 
${\vec \psi}_-$ soliton should be accompanied by oscillations.
The growth rate of the odd perturbation is larger than that of the even one; 
therefore the generic evolution of the instability is expected to be dominated by the odd perturbations. 

On the other hand, for very large
 $a$ --- more precisely,  for 
$\eta_1<\eta<0$, with $\eta_1$ as in \eqref{eta1} --- the spectrum has only one, real, unstable eigenvalue (with an odd eigenvector).
In this range, the instability growth should be initially monotonic.   % and the product(s) of the soliton's decay should break the  spatial parity. 
The moderately large amplitudes (corresponding to $\eta_2<\eta<\eta_1$) are characterised by one or two positive real eigenvalues, 
with odd eigenvector(s), and a complex quadruplet whose eigenvector is even in $X$. 
% Solitons with amplitudes in this range are prone both to a monotonic instability
% with parity-breaking products, and to oscillatory instability preserving the original, even, parity of the soliton. 
The odd eigenfunctions have larger growth rates than the even ones; hence again, the 
monotonically growing odd perturbations
should be dominating the evolution.

Thus, depending on 
 the soliton's amplitude,
the odd perturbations are either the only unstable perturbations of the
${\vec \psi}_-$ soliton, or its dominant unstable perturbations. 
For these,  Eqs.\eqref{P4} give $\delta {\mathcal P}_u= \delta {\mathcal P}_v=0$.
Therefore, the odd perturbations do not  {\it immediately\/} 
induce the asymmetry of the $u$ and $v$ components of the soliton.
On the other hand, 
 the momenta
${\mathcal M}_u$ and ${\mathcal M}_v$ in Eqs.\eqref{M4}
 are both nonzero, with ${\mathcal M}_u \neq {\mathcal M}_v$.
 This means that  the two components, the ``$u$ pulse" and the ``$v$ pulse",  will 
 be set in motion ---  they will start moving with unequal 
velocities  gradually separating in space. 
As a result,
the ``$u$ pulse" will be progressively deprived of the  services of its power-draining partner, 
while  the ``$v$ pulse"  will be cut from the power supply by its active counterpart. 
Eventually,  this will set off the growth of the % $u$-$v$ 
asymmetry in the amplitudes of the pulses --- the ${\mathcal P}_u$ will start growing and ${\mathcal P}_v$ decreasing.

Equations \eqref{M4} can tell us whether the emerging pulses will move in the opposite 
or in the same direction (yet with different velocities). 
Indeed, the fragments will move in the opposite directions if the momenta \eqref{M4} are opposite in signs: $2 (\gamma/\mu)-1<0$.
Recalling that $\mu= \lambda a^2$, this condition can be written as
\be
a^2 > \frac{2 \gamma}{\lambda}.
\label{oppo}
\ee
In the complementary region, $a^2< 2\gamma/\lambda$, the $u$- and $v$-pulses will move in the same direction.

\begin{figure}[t]
 \begin{center} 
 \includegraphics*[width=\linewidth]{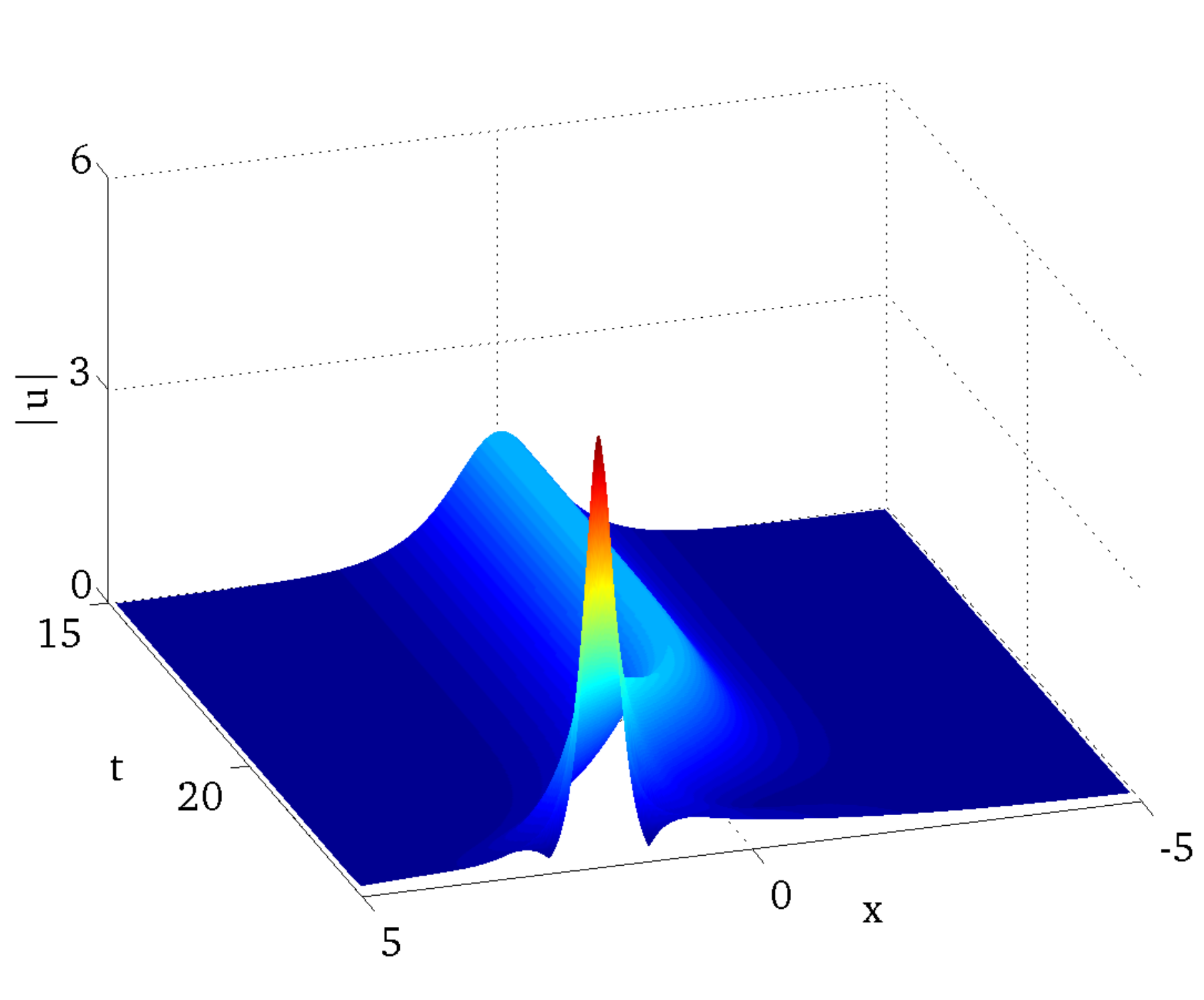} % {fig7a_visible_ridge.eps}
 \includegraphics*[width=\linewidth]{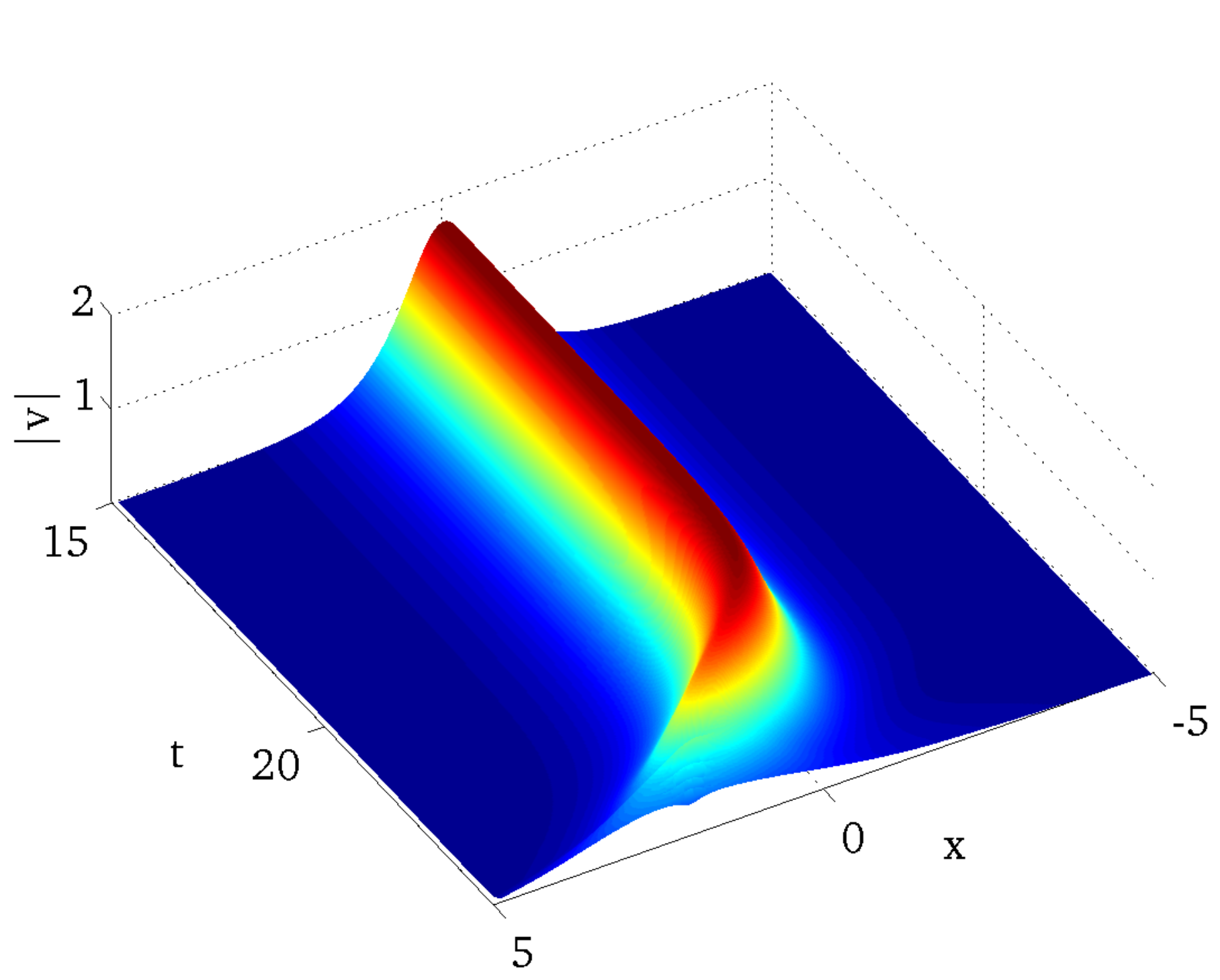} % {fig7b_short_t.eps}   
 \caption{A spontaneous motion  of the  soliton ${\vec \psi}_-$ followed by the blowup
  of its $u$-component and decay of $v$.
  Here $\gamma=0.95$ and $a=1.767$ (${\eta}=- 0.2$).  
 \label{motion_g095_eta02}}
 \end{center}
 \end{figure}

 To verify these predictions, we carried out
numerical simulations of solitons with a range of amplitudes, for small and large values of the 
gain-loss rate
($\gamma=0.1$, $0.2$, $0.5$,  $0.7$, and $0.95$).

In agreement with the linear analysis, the
  small-amplitude ($|\eta|>|\eta_2|$) 
  unstable solitons were detected either to disperse
(a process accompanied by the oscillation of the soliton profiles, see
fig.\ref{decay_or_breather}(a)) or form  long-lived oscillatory states  (fig.\ref{decay_or_breather}(b)).

The instability of solitons with large and moderately large amplitudes ($|\eta|<|\eta_2|$) was seen to grow
monotonically,
at least at the initial stage.
One of the recorded scenarios  starts with  a spontaneous motion of the two components of the soliton
 in the same direction, with 
slightly different velocities (fig.\ref{motion_g095_eta02}), followed
by the 
blowup of the $u$-component and decay of its $v$ counterpart.
This behaviour was detected for the amplitudes $a$ lying outside the region \eqref{oppo}.
(In particular, the unstable soliton shown in fig.\ref{motion_g095_eta02} has $a^2=3.12$ which is smaller than
the value $2 \gamma/\lambda= 4.00$ corresponding to $\gamma=0.95$ and $\eta= -0.2$.)

\begin{figure}[t]
 \begin{center} 
  \includegraphics*[width=\linewidth]{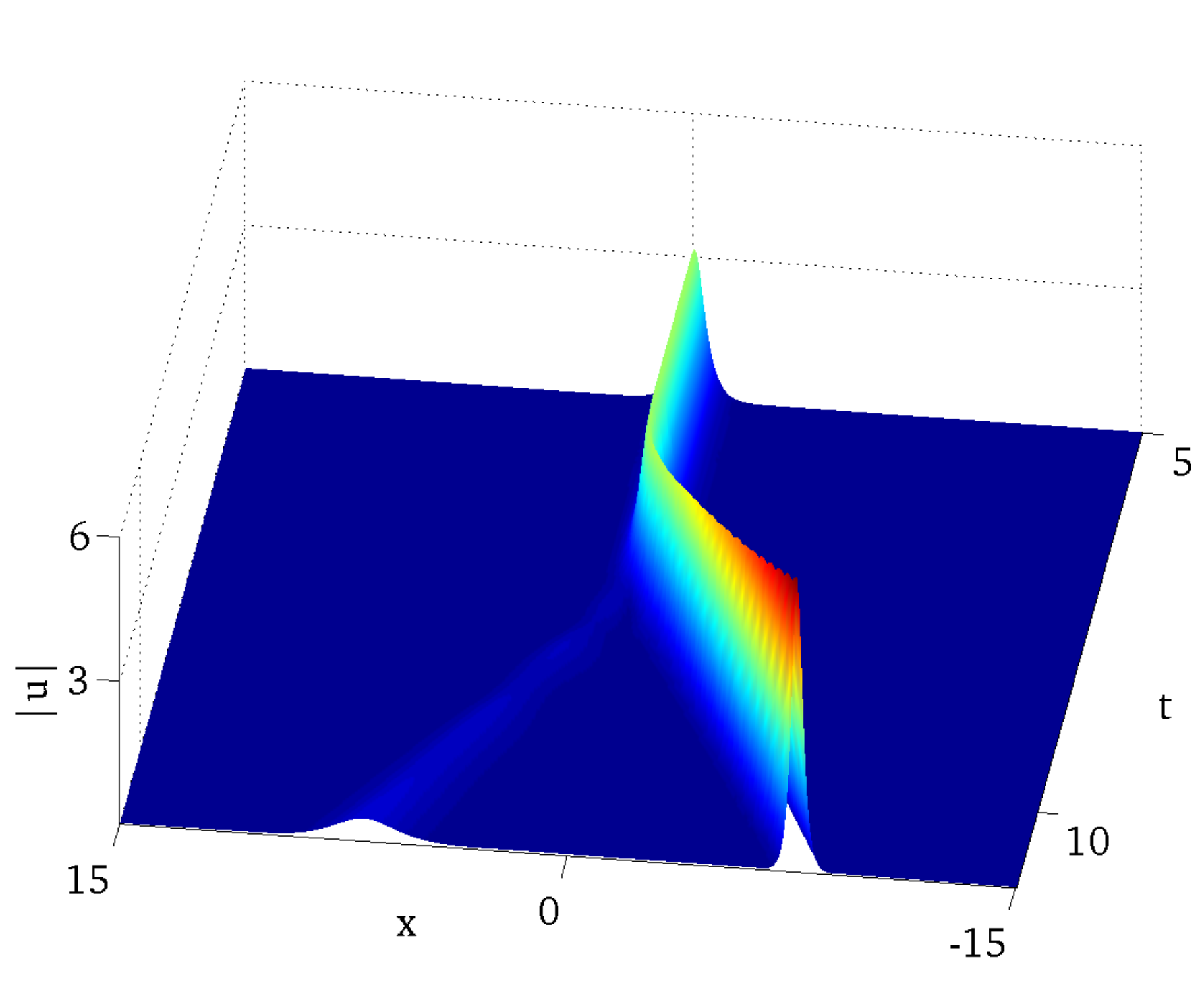} % {fig8a_newt.eps}
 \includegraphics*[width=\linewidth]{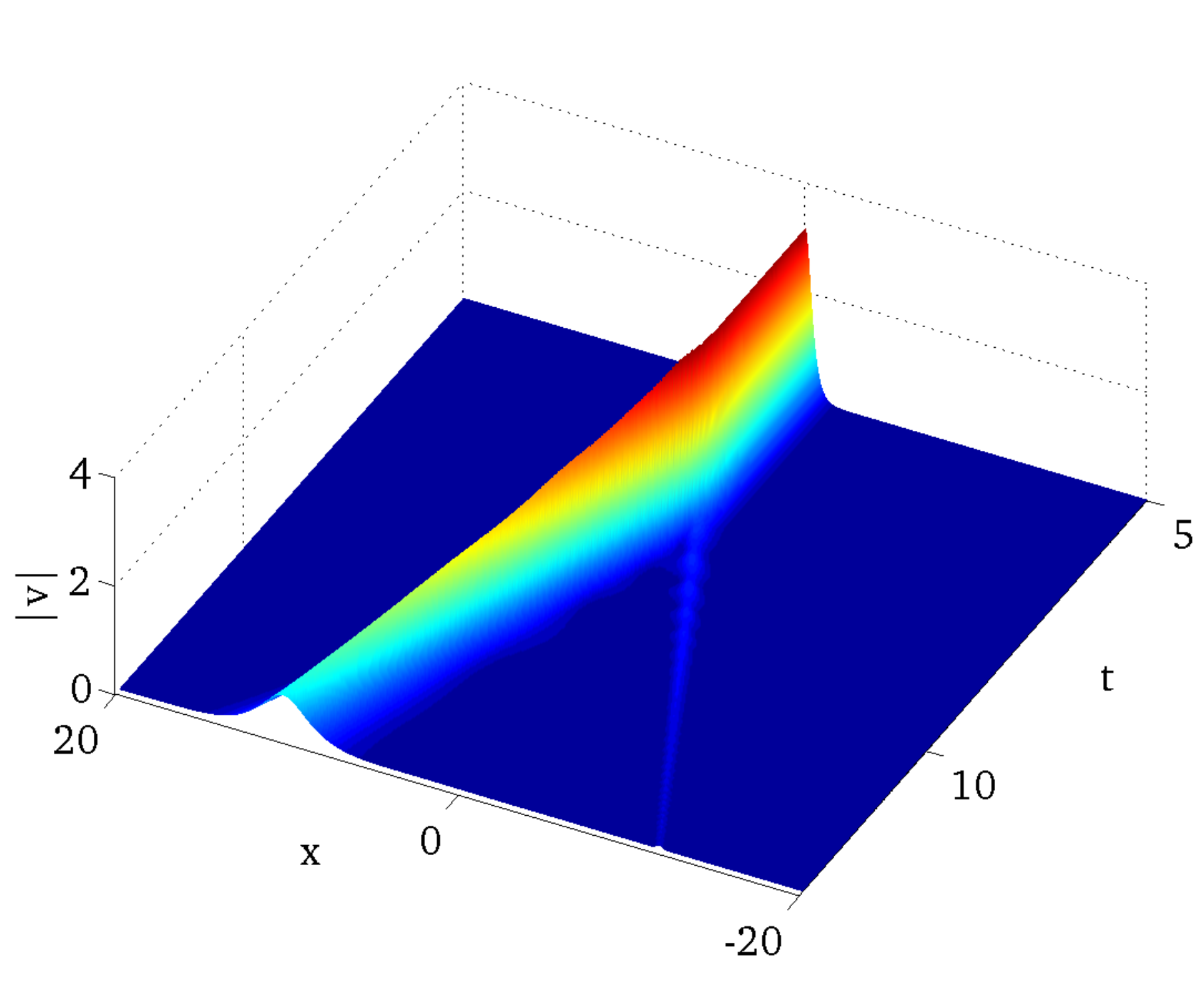} % {fig8b_newt.eps}
     \caption{A dissociation of the  soliton ${\vec \psi}_-$ into a pair of pulses followed by the blowup
  of the pulse with large $u$-component.
  Shown is the magnitude of the field $u$ (a) and field $v$ (b). 
  Here $\gamma=0.1$ and $a=3.15$ (${\eta}=-0.2$).   \label{blow_g01_E_Minus_02}}
 \end{center}
 \end{figure}

\begin{figure}[t]
 \begin{center} 
   \includegraphics*[width=\linewidth]{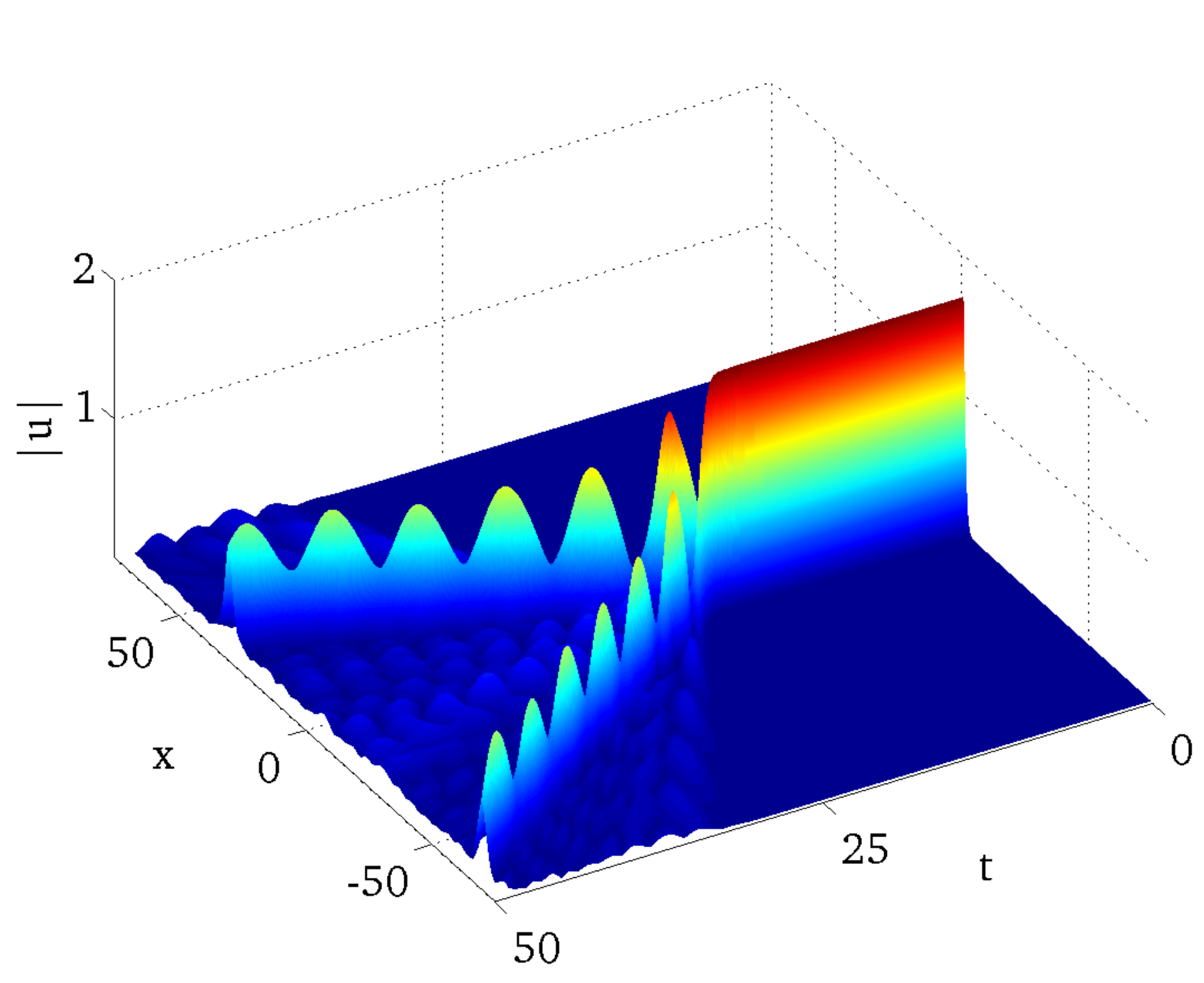} % {double_breathe_new_u.eps}
    \includegraphics*[width=\linewidth]{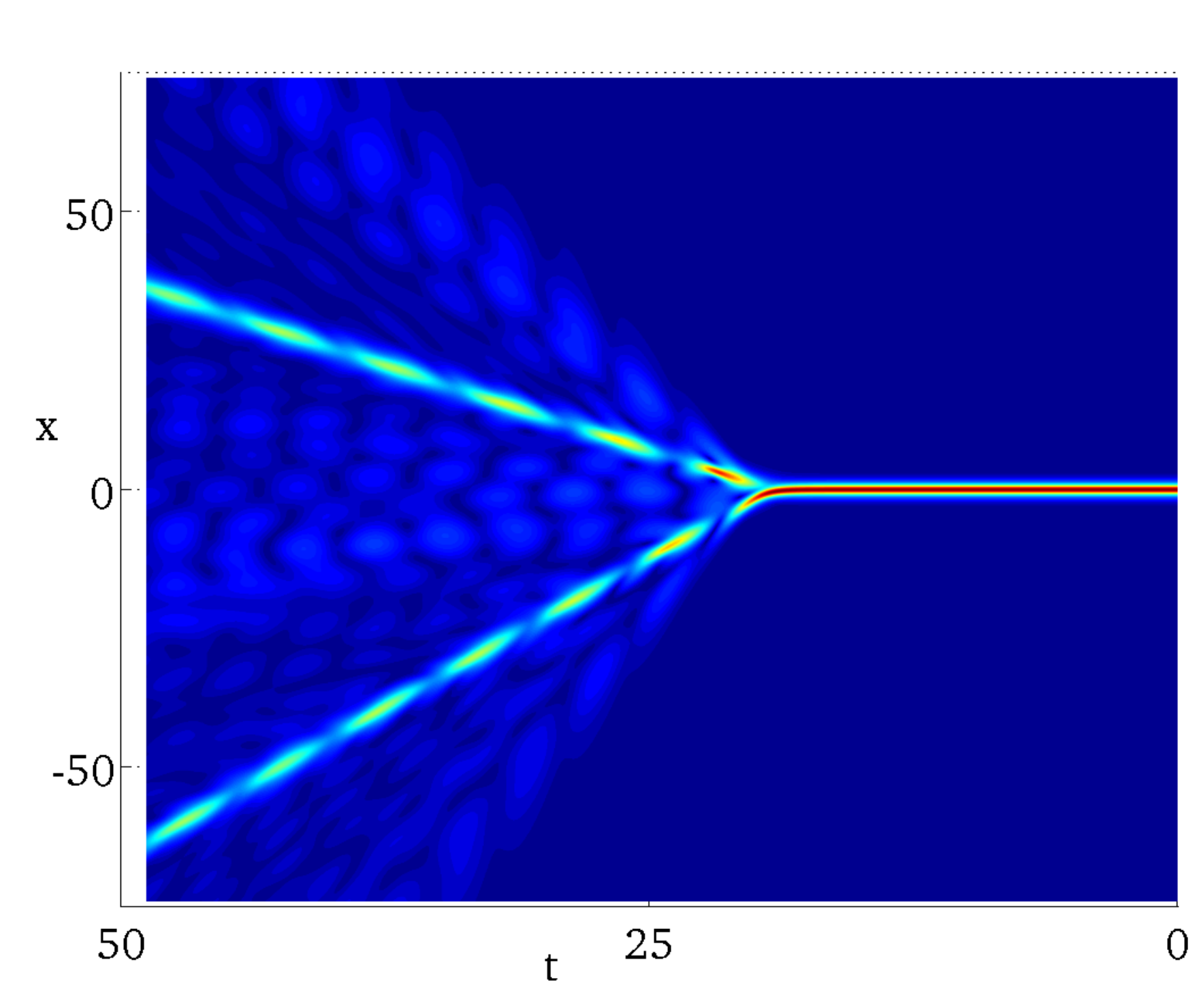} % {fig9b_new2.eps} 
  \caption{A dissociation of the  soliton ${\vec \psi}_-$ into a pair of breathers.
  Shown is the magnitude of  $u$ (a) and  contour plot of $|v|$ (b).
  Here $\gamma=0.1$ and $a=1.686$ (${\eta}=-0.7$).  
%   {\bf (Nora: Change the $v$ panel into a contour map)}.
 \label{breathers_g01_E_Minus_07}}
 \end{center}
 \end{figure}

\begin{figure}[t]
 \begin{center} 
        \includegraphics*[width=\linewidth]{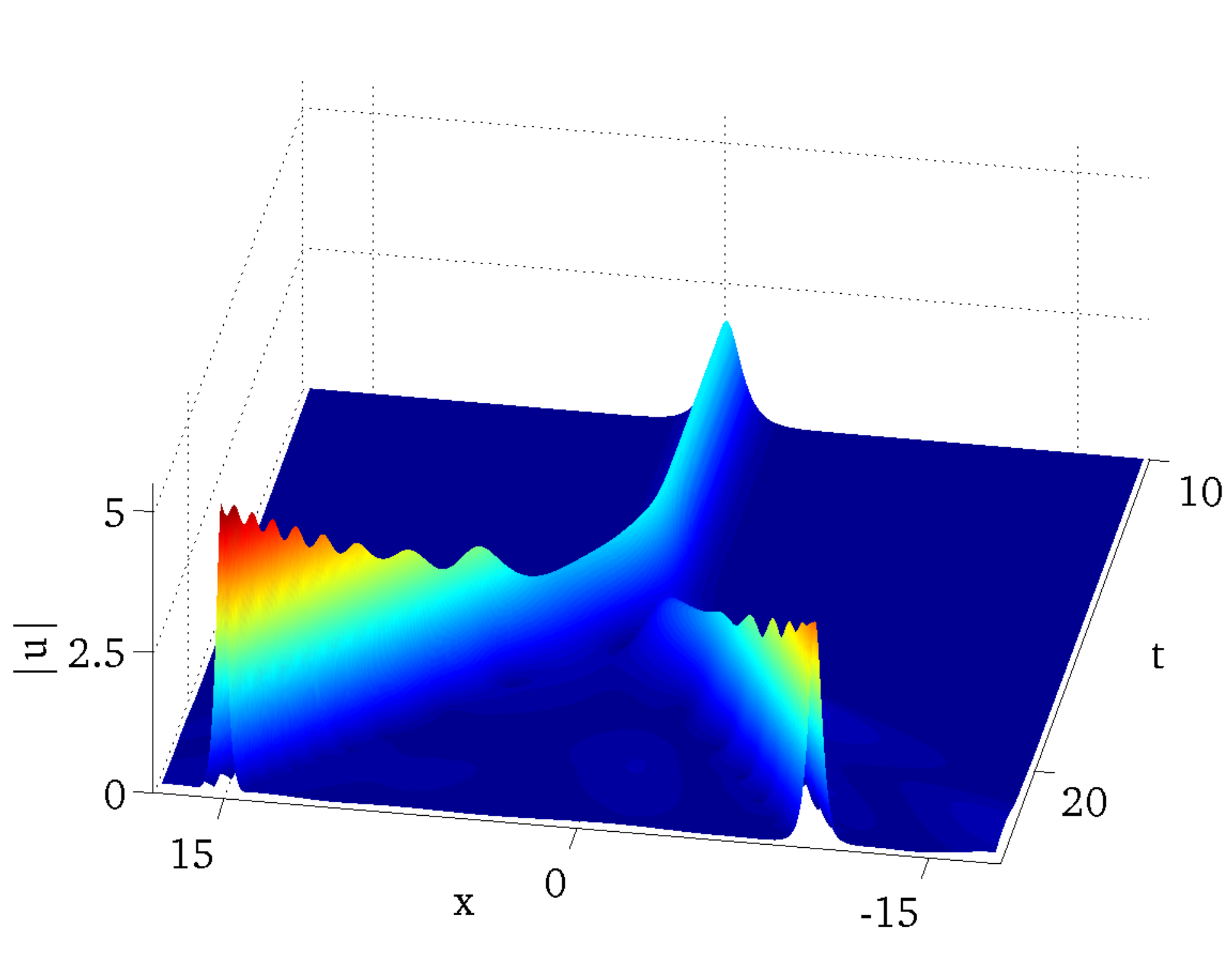} % {fig10a_new3.eps}      
    \includegraphics*[width=\linewidth]{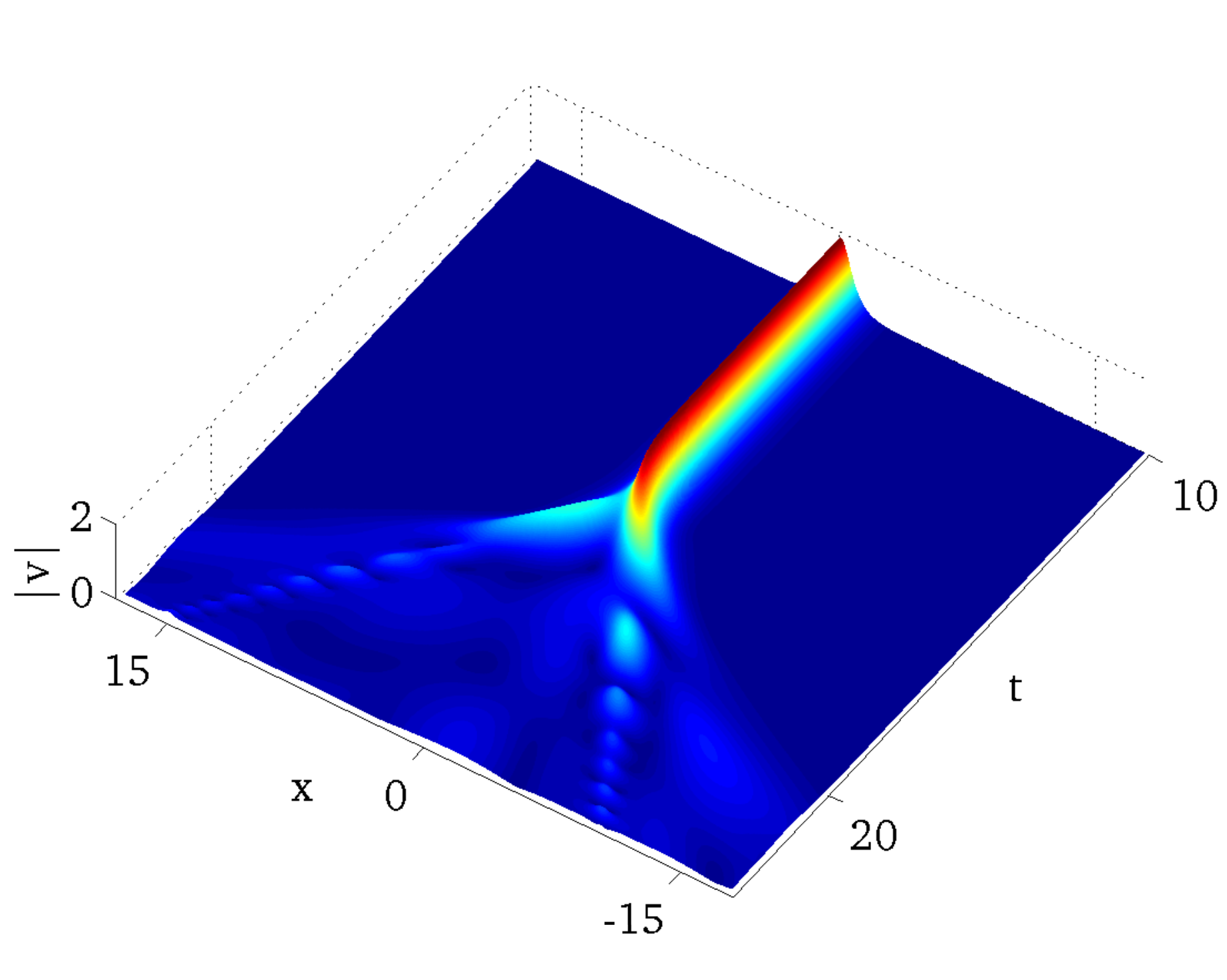} % {fig10b_new3.eps}  
  \caption{A dissociation of the  soliton ${\vec \psi}_-$ into a pair of 
  pulses followed by their blowup.  Shown is the magnitude of  $u$ (a) and  modulus of $v$ (b).
  Here $\gamma=0.1$ and $a=1.8212$ 
($\eta=-0.6$). 
%   {\bf (Nora: Change the $v$ panel into a contour map)}.
 \label{double_blow}}
 \end{center}
 \end{figure}

The other observed evolution starts with  the motion of the two components  in  opposite directions. 
In this case the dissociation of the ${\vec \psi}_-$ soliton may be followed by a blowup of $u$ and decay of $v$
(fig.\ref{blow_g01_E_Minus_02}) or  a formation of a pair of breathers (fig.\ref{breathers_g01_E_Minus_07}). 
Lastly, the linear instability may ``just miss"  the breathers'  basin of attraction in which
case the nuclei of the two breathers will have their $u$ component blow up (fig.\ref{double_blow}). 
These types of evolution were recorded for the amplitudes $a$ satisfying the condition \eqref{oppo}. 
(In particular, fig.\ref{blow_g01_E_Minus_02} corresponds to $a^2=9.92$ and $2 \gamma/\lambda=0.42$;
fig.\ref{breathers_g01_E_Minus_07} to $a^2= 2.84$ and $2 \gamma/\lambda=0.31$; and
fig.\ref{double_blow} to $a^2=3.32$ and $2 \gamma/\lambda= 0.31$.)

\begin{figure}[t]
 \begin{center} 
   \includegraphics*[width=\linewidth]{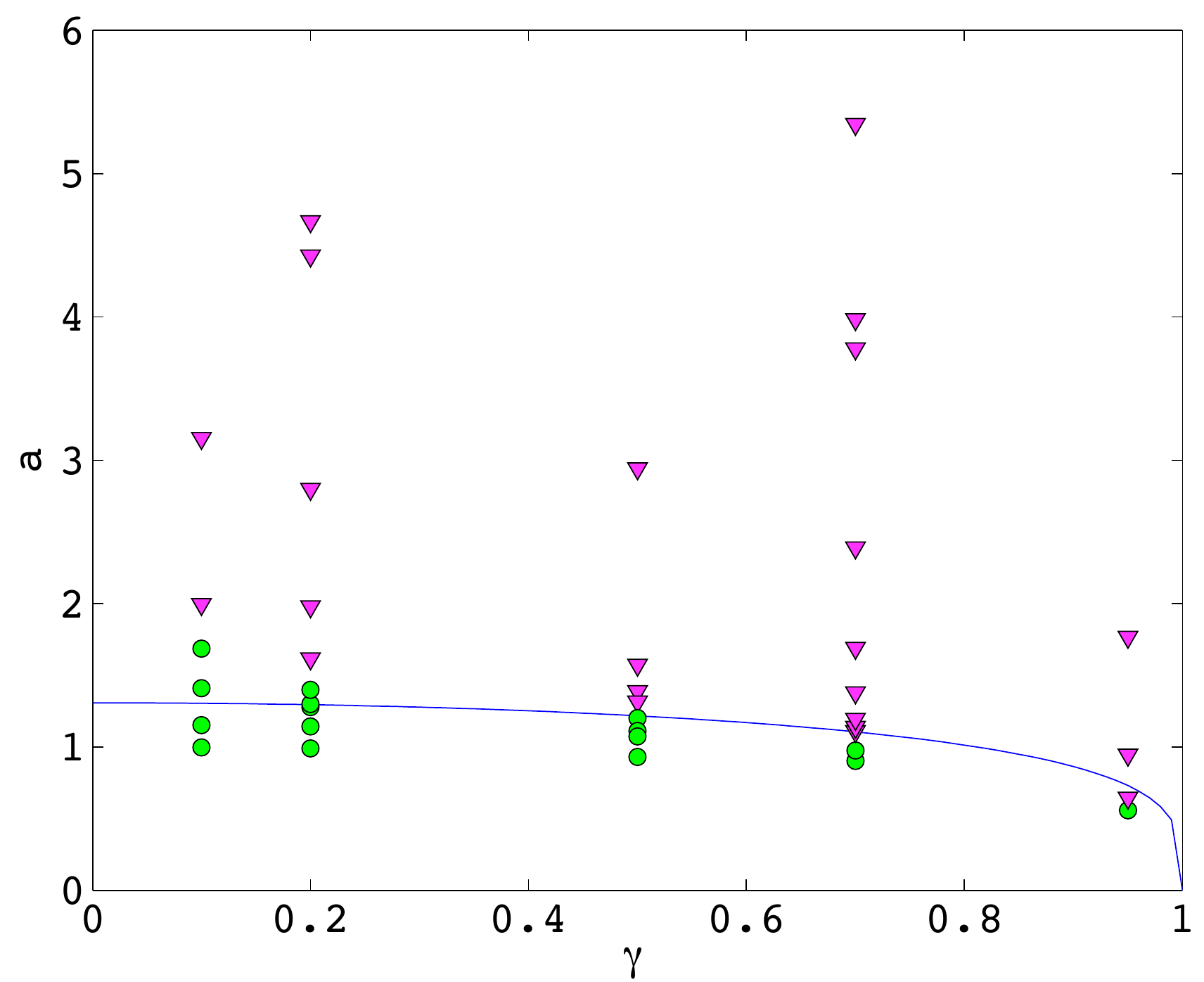} % {new_psiminus_chart.eps}
       \caption{
 Chart of asymptotic regimes emerging from the unstable soliton ${\vec \psi}_-$.
 The  solid curve is given by   $a^2= (-2/\eta_2) \sqrt{1-\gamma^2}$.
  The magenta triangles mark blowups while green circles indicate the bounded
 regimes: dispersion of the soliton or formation of breathers.
  \label{chart_minus}}
  \end{center}
 \end{figure}

Our  numerical simulations of  the unstable regimes are summarised in fig.\ref{chart_minus}.
Triangles mark parameter values for which the unstable soliton or  the two fragments of its break-up were observed to blow up;
circles indicate simulations that ended in the formation of one or two breathers.
The  solid curve in the figure is the demarcation line between the small- and  large-amplitude ranges identified 
in the linear analysis;
the curve is defined by $\eta=\eta_2$ with $\eta_2$ as in \eqref{eta2}.
The chart shows a  clear correlation between 
the type of the 
unstable eigenvalues (real vs complex) 
and
the soliton decay product (blowup vs breathers).

\section{Conclusions}
\label{Conclusions}

We have examined stability of two families of solitons  in the  two-dimensional  $\mathcal{PT}$ symmetric
coupler with gain and loss.  The  dynamical regimes set off by the instability have also been explored.
The results of our study can be summarised as follows.

 1. Despite the presence of
 gain and loss,  the bifurcations occurring 
  in the $\mathcal{PT}$-symmetric
 system \eqref{A1} are of conservative type.
 (Linearised eigenvalues cross from the imaginary to the real axis, or 
 collide, pairwise, on the imaginary axis and emerge into the complex plane.)
 As a result, the soliton instability cannot give rise to localised limit cycles 
 (which would be a typical outcome of the Hopf bifurcation in a dissipative system).
 In the $\mathcal{PT}$-symmetric system,   the soliton instability either triggers its blowup (the process where the amplitude grows
 without bound at an exponential rate) or produces finite-lifetime breathers.

 2. The soliton stability and internal dynamics are  determined 
by a single self-similar combination ($\eta$) of the gain-loss coefficient $\gamma$ and the soliton's amplitude $a$.

3. The high-frequency solitons with amplitudes smaller than $a_c$, where $a_c$ is as in \eqref{A14}, are stable,
and with amplitudes greater than $a_c$, unstable. {\it All\/}  low-frequency solitons are unstable; however the lifetimes of 
the solitons with small amplitudes are exponentially long so for all practical purposes  they can be regarded as stable.

4.  The mechanisms of instability of the high- and low-frequency soliton are different. 
The unstable perturbation of the high-frequency soliton triggers the growth of asymmetry between
the active ($u$) and lossy ($v$) components of the soliton, destroying the gain/loss balance in the system.
The unstable perturbation of the low-frequency soliton also upsets the energy balance;  however this time
it is done by splitting the $u$ and $v$ components off from their common axis. 
The difference in the instability mechanisms is reflected in the difference in the products of the 
high-
and low-frequency soliton  breakup.

\section*{Acknowledgements}

NA and IB's work in Canberra was funded via Visiting Fellowships of the 
Australian National University; also supported by the National Research Foundation 
of South Africa
(grants UID 67982 and 78952).
This project was carried out with the assistance of the Australian Research Council which included the
Future Fellowship FT100100160.
Computations were performed using facilities provided by the University
of Cape Town's ICTS High Performance Computing team.
\appendix

\section{Real-imaginary eigenvalue transitions}
\label{RIT}

The aim of this Appendix is to describe the transitions of the pair of eigenvalues of the 
operator \eqref{A11} from the real to imaginary axis and vice versa, as the parameter $\eta$ crosses
through zero in either direction. 

The numerical analysis indicates  that when $\eta$  approaches 0,
the eigenvalue scales as $\eta^{1/2}$. This suggests an expansion of the form
\begin{align}
\lambda= \eta^{1/2} \lambda_1 + \eta \lambda_2 + ...; \nonumber \\
g= g_0+ \eta^{1/2} g_1 + \eta g_2 +...,  \nonumber  \\
f= f_0 + \eta^{1/2} f_1 + \eta f_2+ ... .
\label{Y1}
\end{align}
This expansion will be validated if all coefficient functions are
found to be bounded and decaying to zero as $X \to \pm \infty$.

The expansion \eqref{Y1}  is similar to the one appearing in the parametrically driven damped nonlinear Schr\"odinger equation  \cite{Barashenkov:1991-113:EPL, Barashenkov:2002-104101:PRL}.
The difference of Eq.\eqref{A11} from the eigenvalue problem in \cite{Barashenkov:1991-113:EPL, Barashenkov:2002-104101:PRL}, is that 
the parametric driving breaks only {\it one\/} of the invariances of the nonlinear Schr\"odinger 
(only the operator $L_0$ is perturbed). On the other hand, 
in Eq.\eqref{A11}, both $L_0$ and $L_1$ acquire perturbations. The 
consequence of this will be the motion of {\it two\/} pairs of eigenvalues through the origin on the $\lambda$ plane.
While one pair is moving from the real to imaginary axis,  the other pair will be moving in the opposite direction.

Substituting \eqref{Y1}  in \eqref{A11} and equating coefficients of like powers of $\eta$, 
we obtain a chain of equations for $\lambda_n$, $g_n$ and $f_n$. The order $\eta^0$ gives
\[
L_0 f_0=0,   \quad
L_1 g_0=0. 
\]
These two equations  coincide with equations for the $U(1)$ and translational
zero modes of the scalar cubic nonlinear Schr\"odinger  equation. 
The bounded solutions are
\[
f_0= C_1 \sech X, \quad g_0= C_2 \sech X \tanh X,
\]
where $C_1$ and $C_2$ are arbitrary constants. 

At the next, $\eta^{1/2}$, order, we get a pair of nonhomogeneous equations
\[
L_0 f_1= \lambda_1 g_0, \quad
L_1 g_1= -\lambda_1 f_0,
\]
whose solutions are
\[
f_1= \frac{C_2}{2} \lambda_1 X  \sech X,
\quad
g_1= \frac{C_1}{2}  \lambda_1    \sech X  (1- X \tanh X).  \label{Y2}
\]
In the context of  stability of the scalar cubic nonlinear Schr\"odinger soliton,
the generalised eigenvectors $(0,f_1)^T$ and $(g_1,0)^T$ would
correspond to the Galilean invariance of that equation and its soliton frequency variations.

Finally,  the order $\eta^1$ gives
\begin{align}
L_0 f_2= -f_0+ \lambda_1 g_1+ \lambda_2 g_0,
\label{Y3} \\
L_1 g_2=-g_0 -\lambda_1 f_1 -\lambda_2 f_0. 
\label{Y4}
\end{align}
The solvability condition for Eq.\eqref{Y3} is 
\be
-\langle z_0| f_0 \rangle + \lambda_1 \langle z_0 | g_1\rangle =0, \label{Y5}
\ee
and the one for \eqref{Y4} is
\be
-\langle y_1 |g_0 \rangle -\lambda_1 \langle y_1 |f_1\rangle=0.
 \label{Y6}
\ee
In \eqref{Y5}-\eqref{Y6}, $z_0=\sech X$ and $y_1= \sech X \tanh X$ 
are 
the null eigenvectors of the operator $L_0$ and $L_1$, respectively. 
The bra-ket notation is used for the $\mathcal{L}^2$ scalar product:
\[
\langle y | z \rangle = \int y(X) z(X) dX.
\]

Substituting for $f_{0,1}$ and $g_{0,1}$ in \eqref{Y5}-\eqref{Y6}, 
the solvability conditions reduce to 
\begin{align*}
C_1 \left(\frac{\lambda_1^2}{2} - 2 \right)=0, \\
C_2 \left(\frac{\lambda_1^2}{2} +\frac23 \right)=0,
\end{align*}
whence either $\lambda_1=\pm 2$ and $C_2=0$, or $\lambda_1= \pm 2i/ \sqrt{3}$ and $C_1=0$. 

This gives us the the leading-order expressions for the two pairs of eigenvalues and eigenvectors
of the operator \eqref{A11} with small $\eta$.
One pair of eigenvalues is
$\lambda = \pm 2 \sqrt{\eta}+O(\eta)$; it is associated with even eigenfunctions:
\[
\left( \begin{array}{c} g \\ f \end{array} \right)=
\left( \begin{array}{c} \pm \sqrt{\eta}              \sech X                (1-X \tanh X)  \\  \sech X \end{array} \right)+ O(\eta).
\]
This pair moves from the real to imaginary axis as $\eta$ decreases from positive to negative values.

The other pair of eigenvalues  is  $\lambda= \pm 2 \sqrt{-\eta/3} +O(\eta)$. The corresponding eigenfunctions are
odd:
\[
\left( \begin{array}{c} g \\ f \end{array} \right)=
\left( \begin{array}{c}  \sech X \tanh X \\ \pm  i \sqrt{\frac{\eta}{3}} X \sech X   \end{array} \right)+ O(\eta).
\]
As $\eta$ moves  from positive to negative, this pair of eigenvalues  translates from the imaginary to real axis.

\section{Asymptotic eigenvalues as $|{\eta}| \to \infty$}
\label{asympt}

In this Appendix, we determine the asymptotic behaviour of 
eigenvalues of the operator \eqref{A11} as ${\eta}$ tends to $\pm \infty$.

We expand
\begin{align*}
f= f_0+ \frac{f_1}{{\eta}} + \frac{f_2}{{\eta}^2}+... , \\
g= g_0+ \frac{g_1}{{\eta}} + \frac{g_2}{{\eta}^2}+... ; 
\\
% \lambda= i\left({\eta} +  \lambda_0 + \frac{\lambda_1}{{\eta}} +  \frac{\lambda_2}{{\eta}^2} +...\right).
\lambda= i\left(\lambda_{-1}  {\eta} +  \lambda_0 + \frac{\lambda_1}{{\eta}} +  \frac{\lambda_2}{{\eta}^2} +...\right).
\end{align*}
It is convenient to introduce coefficient functions $\mathcal{A}_n(X)$ and $\mathcal{B}_n(X)$, 
\[ 
\mathcal{A}_n= f_n-ig_n, \quad \mathcal{B}_n= -(f_n+ig_n).
\]
Substituting in \eqref{A11},
and equating coefficients of  ${\eta}^{-n}$, gives
\be
\lambda_{-1}^2=1, \quad \mathcal{A}_0=0 \label{X1}
\ee
at the order ${\eta}^1$, and
\be
(L_{1/2}-  \lambda_0) \mathcal{B}_n = 2 \sech^2X \mathcal{A}_n+ \sum_{m=1}^n \lambda_m \mathcal{B}_{n-m}  \label{A15}
\ee
\be
2 \mathcal{A}_{n+1}=         2 \sech^2 X \mathcal{B}_n - (L_{1/2}+ \lambda_0) \mathcal{A}_n 
-\sum_{m=1}^n \lambda_m \mathcal{A}_{n-m}   \label{A16}
\ee
for all $n \geq 0$. Here we have introduced an operator 
\[
L_{1/2} = - d^2/dX^2 + 1 - 4 \sech^2 X.
\]

In Eq.\eqref{X1} we can choose,
without loss of generality,
 $\lambda_{-1}=1$. [The other root, $\lambda_{-1}=-1$, defines
the opposite eigenvalue, $-\lambda$, with the eigenvector $(g, -f)^T$.]

Equation \eqref{A15} with $n=0$ is an eigenvalue problem for the operator $L_{1/2}$:
\[
L_{1/2} \mathcal{B}_0(X)= \lambda_0 \mathcal{B}_0(X).
\]
The potential $-4 \sech^2 X$ is  of the P\"oschl-Teller variety so the eigenvalues can be found exactly.
There are two discrete eigenvalues, $\rho_A$ and $\rho_B$:
\be
\rho_A= \alpha-3 \approx -1.438, \quad \rho_B= 3 \alpha-4  \approx 0.685,
\label{rhoAB}
\ee
where $\alpha=\frac12 (\sqrt{17}-1)$.
(The corresponding 
eigenfunctions are  $y_A= \sech^\alpha X$ and $y_B=\sech^{\alpha-1} X \tanh X$,
respectively.)
Thus the correction $\lambda_0$ may take
either of these two real values, $\rho_A$ or $\rho_B$; depending on the choice, the function 
$\mathcal{B}_0(X)$ is even or odd. In either case, $\mathcal{B}_0$ can be chosen real.

Once $\lambda_{-1}$, $\lambda_0$  and $\mathcal{B}_0$ have been chosen, all higher-order 
coefficients  $\mathcal{A}_n, \mathcal{B}_n$ and $\lambda_n$ are determined uniquely.
Furthermore, one can prove by induction that  all these coefficients  are real.
The proof proceeds in three steps.

First, we assume that  $\mathcal{B}_0, \mathcal{B}_1, ..., \mathcal{B}_\ell$, $\mathcal{A}_0, \mathcal{A}_1, ..., \mathcal{A}_\ell$,  
and $\lambda_0, \lambda_1, ..., \lambda_\ell$ with some $\ell \geq 0$
have been found and  are all real. 
Then equation \eqref{A16} gives us  $\mathcal{A}_{\ell+1}$, which  does not have an imaginary component either.

Next we turn to the equation \eqref{A15} with $n=\ell+1$. The solvability condition 
for $\mathcal{B}_n$ is
\begin{align}
\lambda_{\ell+1} \langle \mathcal{B}_0 | \mathcal{B}_0 \rangle =-  \sum_{m=1}^{\ell} \lambda_m  \langle \mathcal{B}_0|\mathcal{B}_{\ell+1-m} \rangle \nonumber
\\
- 2 \int \mathcal{B}_0     \sech^2 X  \mathcal{A}_{\ell+1}       dX.
\end{align}
This equation expresses $\lambda_{\ell+1}$ through  $\mathcal{A}_{\ell+1}$,   $\mathcal{B}_1, \mathcal{B}_2, ..., \mathcal{B}_{\ell}$, and $\lambda_1, \lambda_2, ..., \lambda_{\ell}$.
By our assumption, all these coefficients are real; hence $\lambda_{\ell+1}$ does not have an imaginary part either.
Now that the solvability condition has been satisfied, the nonhomogeneous equation  \eqref{A15} can be solved for $\mathcal{B}_{\ell+1}$.
The right-hand side 
in \eqref{A15}  includes real  $\mathcal{A}_{\ell+1}$,  $\mathcal{B}_0, \mathcal{B}_1, ..., \mathcal{B}_{\ell}$,  
and $\lambda_1, \lambda_2, ..., \lambda_{\ell+1}$.
Therefore, the solution $\mathcal{B}_{\ell+1}(X)$ is  real as well. This completes the proof.

Note that if we choose $\lambda_0=\rho_A$, all the coefficient functions $\mathcal{B}_n(X)$ and $\mathcal{A}_n(X)$
are even, whereas if we set $\lambda_0=\rho_B$, all functions are odd.

Thus we conclude that in the limit $|{\eta}| \to \infty$, the real part of the eigenvalue in the problem \eqref{A11} 
is zero to all orders in ${\eta}^{-n}$. This means that the real part is either
exactly zero or exponentially small in $|{\eta}|$: $\mathrm{Re} \, \lambda \sim e^{-\alpha |{\eta}|}$, $\alpha>0$.
The imaginary part may assume one of the two pairs of values,
$\mathrm{Im} \, \lambda_A= \pm [{\eta}+  \rho_A +O({\eta}^{-1})]$ or $\mathrm{Im} \, \lambda_B= \pm [{\eta}+  \rho_B +O({\eta}^{-1})]$,
where $\rho_A<0$ and $\rho_B>0$ are given by Eq.\eqref{rhoAB}. The eigenfunctions $f(X)$ and $g(X)$
associated with  the former are even and those pertaining to the latter are odd.

\section{Real eigenvalue in $-1<{\eta}<0$}
\label{E-1}

In this Appendix, we prove the existence of real 
eigenvalues of the symplectic operator \eqref{Z0} with negative ${\eta}$,
and discuss their behaviour as $\eta \to -1$.

If ${\eta}<0$,      neither of the scalar operators in \eqref{A11}
is invertible on the full $\mathcal{L}^2$ space.
Fortunately,
 the symplectic operator \eqref{Z0} is parity preserving and so all its eigenfunctions 
fall into one of the two broad classes: even and odd ones. 
Therefore, when considering eigenvalues $\lambda$ with even eigenfunctions, 
we can restrict the scalar operators to the subspace of  $\mathcal{L}^2$ consisting of even functions.
When examining the odd eigenfunctions, these operators can be restricted to the odd subspace.

The advantage of the separate treatment of eigenfunctions with different parity becomes clear when we note that 
the operator $L_0+ {\eta}$ with $-1<{\eta} \leq 0$ is 
positive definite on the   subspace of odd functions --- denoted $\frak{S}$ in what follows.
Therefore, if we confine ourselves to eigenvalues $-\lambda^2$ of the generalised eigenvalue problem \eqref{A12}
associated with odd eigenfunctions $g(X)$, 
the lowest of these  ``odd" eigenvalues will be given by
 the minimum of 
the Rayleigh quotient \eqref{A13} on $\frak{S}$:
\be
-\lambda^2 = \min_{g \in \frak{S}} \frac{\langle g |L_1+{\eta}|g \rangle}{\langle g |(L_0+{\eta})^{-1}|g \rangle}.
\label{A26}
\ee

The eigenfunction $y_1= \sech X \tanh X$ associated with the negative eigenvalue ${\eta}$ of the operator $L_1+{\eta}$, is odd. Hence  
 the quotient in \eqref{A26} can assume negative values in  $\frak{S}$, and its minimum $-\lambda^2$ is negative. 
We conclude that in the parameter region $-1<{\eta} \leq 0$,
  Eq.\eqref{A11} has a pair of opposite real eigenvalues $\pm \lambda$  (with odd eigenfunctions).

 Contrary to what one might have expected, this pair of  eigenvalues does not converge at the origin
as ${\eta}$ reaches $-1$. The following argument shows that the  eigenvalues should
remain finite as ${\eta} \to -1+0$.

First, we note that any function $g(X)$ from $\frak{S}$ can be expanded 
over the complete set of odd eigenfunctions of the operator $L_0$:
\be
g(X)=  \int _0^\infty \mathcal{G}(k) z_k(X) dk.
\label{A70}
\ee
Here
\[
z_k= \frac{\tanh X \cos (kX)+ k \sin (kX)}{\sqrt{\pi (1+k^2)}}
\]
is a continuous spectrum eigenfunction pertaining to the 
eigenvalue $E(k)=1+2k^2$. 
The expansion \eqref{A70} does not include a sum over discrete eigenvalues because $L_0$ has only one such eigenvalue
and the corresponding eigenfunction is even.

Letting ${\eta}=-1+\epsilon$, $\epsilon>0$, and
substituting \eqref{A70} in \eqref{A26},    the denominator can be written as
\be
\langle g |(L_0+{\eta})^{-1}|g \rangle= \int_{-\infty}^\infty  |\mathcal{G}(k)|^2 (2k^2+ \epsilon)^{-1}  dk.
\label{A71}
\ee

Now consider a subspace $\frak{S}^\prime$ of  $\frak{S}$ 
consisting of functions $g$ satisfying two conditions.
The first requirement is that 
the function $g(X)$  should decay to zero,  as $X \to \pm  \infty$,  faster than $|X|^{-n}$ with any  $n>0$.
This condition ensures that the coefficient function $\mathcal{G}(k)$ has derivatives of all orders at $k=0$
and hence can be expanded in a Taylor series centered on that point.

The second condition is that   $g(X)$ should satisfy
\be
\int g(X) z_{k=0}(X) dX=0,
\label{A30}
\ee
where $z_{k=0}=\frac{1}{\sqrt{\pi}}  \tanh X$ is the eigenfunction pertaining to the edge of the continuous
spectrum of $L_0$.  Owing to the constraint \eqref{A30},  the function $\mathcal{G}(k)$ has a zero at  $k = 0$;
therefore
 its Taylor expansion 
has the form
$\mathcal{G}(k)= \mathcal{G}_1 k + \mathcal{G}_2 k^2+...$. 
Consequently,  $|\mathcal{G}(k)|^2 k^{-2}$ does not have a singularity at $k=0$,
  the integral $\int |\mathcal{G}(k)|^2 k^{-2}  dk$
is convergent and the denominator   \eqref{A71} remains finite as $\epsilon \to 0$.

If we  find a  function $g$ in $\frak{S}^\prime$  which renders the numerator in \eqref{A26} negative, 
we will obtain a simple bound for the eigenvalue $\lambda$ at the point $\eta=-1$:
\be
-\lambda^2 < \frac{\langle g |L_1-1|g \rangle}{\int |\mathcal{G}(k)|^2 k^{-2}  dk}<0.
\label{A31}
\ee

To verify that functions with these properties do exist, take, for instance, 
\[
g(X)= \sech X \tanh X - \frac12 X \sech X.
\]
This odd function decays faster than any power of $X$ and satisfies the constraint \eqref{A30}; hence it is  in $\frak{S}^\prime$.
A simple integration gives
\[
\langle g| (L_1-1) | g \rangle= \frac32- \frac{11}{72} \pi^2,
\]
which is negative  (approximately $-7.9 \times 10^{-3}$).

This completes our argument. We have shown that the positive eigenvalue $\lambda$
of the
symplectic operator \eqref{Z0}, which had been proven to exist for $-1<\eta<0$, does not 
approach zero as $\eta$ approaches $-1$ from the right. (The same is obviously true for the negative counterpart
of this positive $\lambda$.)
Note that our argument does not rule out the existence of pairs of real eigenvalues converging at zero 
as $\eta$ approaches $-1$ from the left.

%-----------------------------------------------------------------
%\bibliographystyle{osa3}
\bibliography{db_Alexeeva_arXiv}
%-----------------------------------------------------------------

\end{document}